\documentclass[10pt,aps,twocolumn,secnumarabic,balancelastpage,amsmath,amssymb,nofootinbib,hyperref=pdftex,prb,citeautoscript,superscriptaddress]{revtex4-1}

\usepackage{graphics}
\usepackage[pdftex]{graphicx}
\usepackage{tabularx}
\graphicspath{{./Figures/}}
\usepackage{longtable}
\usepackage{amsmath}
\usepackage{color,soul}

\renewcommand\thesection{\Roman{section}}

\makeatletter
\def\p@subsection{\thesection\,}
\makeatother

\newcommand{\sgn}[1]{\text{sgn}(#1)}

\newcommand{\kk}{{\bf k}}

\usepackage{hyperref}
\hypersetup{colorlinks=true,linkcolor=blue,anchorcolor=blue,citecolor=blue,filecolor=blue,urlcolor=blue,bookmarksnumbered=true,pdfview=FitB}

\begin{document}

\title{Transport signatures of topological superconductivity in a proximity-coupled nanowire}
\author{Christopher Reeg}
\affiliation{Department of Physics, University of Basel, Klingelbergstrasse 82, CH-4056 Basel, Switzerland}
\author{Dmitrii L. Maslov}
\affiliation{Department of Physics, University of Florida, P. O. Box 118440, Gainesville, FL 32611-8440, USA}
\date{\today}
\begin{abstract}
We study the conductance of a junction between the normal and superconducting segments of a nanowire, both of which are subjected to spin-orbit coupling and an external magnetic field. We directly compare the transport properties of the nanowire assuming two different models for the superconducting segment: one where we put superconductivity by hand into the wire, and one where superconductivity is induced through a tunneling junction with a bulk $s$-wave superconductor. While these two models are equivalent at low energies and at weak coupling between the nanowire and the superconductor, we show that there are several interesting qualitative differences away from these two limits. In particular, the tunneling model introduces an additional conductance peak at the energy corresponding to the bulk gap of the parent superconductor. By employing a combination of analytical methods at zero temperature and numerical methods at finite temperature, we show that the tunneling model of the proximity effect reproduces many more of the qualitative features that are seen experimentally in such a nanowire system.
\end{abstract}

\maketitle

\section{Introduction}
\begin{figure}[t!]
\includegraphics[width=\linewidth]{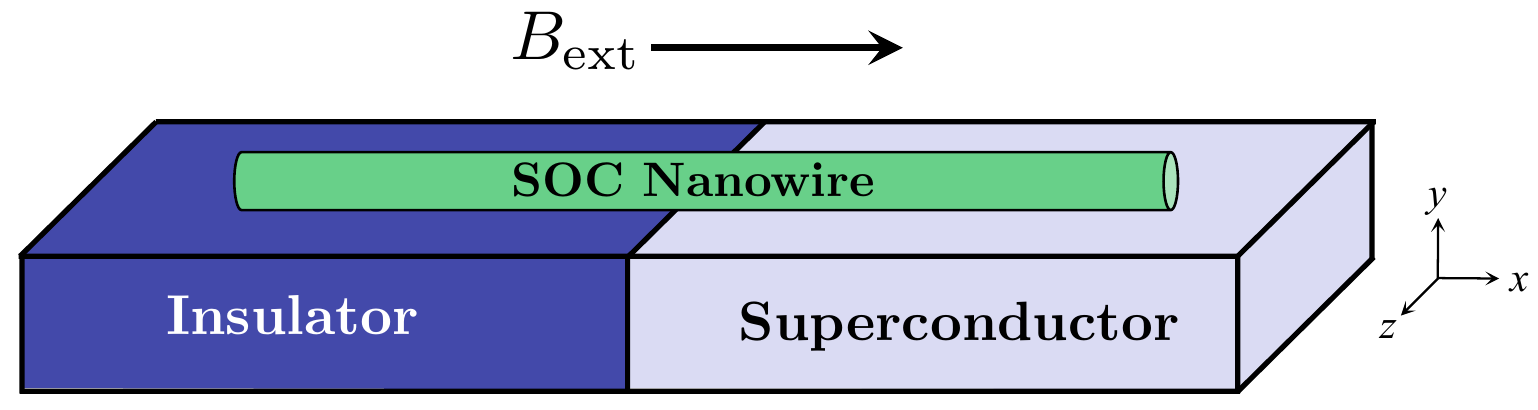}
\caption{\label{geometry}Model geometry. A 1D nanowire (of infinite length) with spin-orbit coupling is placed on top of a junction between an insulator ($x<0$) and a superconductor ($x>0$). A magnetic field ${\bf B}_\text{ext}$ is applied perpendicularly to the effective Rashba field. We chose the latter to be along the $z$-axis and ${\bf B}_\text{ext}$ to be along the wire.}
\end{figure}

There has been an intense effort in recent years to realize zero-energy Majorana bound states in condensed matter systems \cite{Alicea:2012,Leijnse:2012,Beenakker:2013} due to potential applications of such states in quantum computing. \cite{Kitaev:2001,Nayak:2008} The most promising proposal to date for engineering topological superconductivity involves applying a magnetic field
to a nanowire containing both spin-orbit-coupling (SOC) and proximity-induced superconductivity, \cite{Oreg:2010,Lutchyn:2010} as shown in Fig.~\ref{geometry}. By increasing the magnetic field strength in such a setup, one tunes through a topological phase transition. In the topological phase, the zero-energy Majorana mode that is localized at the end of the superconducting segment of the nanowire is expected to produce a quantized (at $2e^2/h$) zero-bias peak in the differential conductance through the normal/superconducting junction at zero temperature. \cite{Sengupta:2001,Law:2009,Flensberg:2010,Wimmer:2011,Fidkowski:2012} Similarly, there have also been related proposals to detect the Majorana modes by coupling a quantum dot to the superconducting segment rather than a normal lead. \cite{Flensberg:2011,Leijnse:2011,Liu:2011,Vernek:2014,Prada:2017} Because of the abundance and relative simplicity of the required ingredients, these theoretical proposals very quickly received a great deal of experimental attention. \cite{Mourik:2012,Deng:2012,Das:2012,Finck:2013} Since the first generation of experiments, great progress has been made both in the fabrication of cleaner devices and in the improvement of the induced superconductivity. \cite{Chang:2015,Albrecht:2016,Kjaergaard:2016,Zhang:2016,Kjaergaard:2017,Deng:2016} However, conclusive evidence for topological superconductivity remains elusive.

The experimental progress has in turn motivated much theoretical investigation aimed at describing the conductance of such a nanowire system. Most of these theoretical studies focus on numerical simulations of complex geometries in an attempt to quantitatively reproduce the experimentally measured conductance, \cite{Stanescu:2011,Prada:2012,Lin:2012,Pientka:2012,Gibertini:2012,Rainis:2013,Roy:2013prb,Roy:2013,Stanescu:2014,Liu:2017} while there have been far fewer analytical studies. \cite{Qu:2011,Yan:2014,Setiawan:2015} Additionally, nearly all previous transport models (with the exception of the numerical models of Refs.~\onlinecite{Stanescu:2011,Stanescu:2014}) do not account for the fact that electrons in the wire can spend a fraction of their time in the proximity-coupled superconductor. Instead, the proximity effect is most frequently modeled by the presence of an intrinsic pairing mechanism in the nanowire, while the parent superconductor is neglected completely. However, this is a reasonable approximation only in the limit of low energies and weak tunneling between superconductor and nanowire (as we will show explicitly). There has been significant discussion of the importance of treating the parent superconductor explicitly when analyzing the proximity effect theoretically, as the proximity coupling can strongly renormalize the bulk properties of the nanowire. \cite{Sau:2010prox,Potter:2011,Stanescu:2011,Takei:2013,Peng:2015,Cole:2015,Hui:2015,Cole:2016,vanHeck:2016,Stanescu:2017} As the experimentally achievable coupling strength continues to increase, it becomes more important to understand the effect of the parent superconductor on the transport properties of the nanowire as well.

In this paper, we calculate the conductance spectrum in the nanowire geometry within the Blonder-Tinkham-Klapwijk (BTK) theory \cite{Blonder:1982} using two different models for the proximity effect between the underlying superconductor and the wire to which it is coupled. 

First, we model the proximity effect through the inclusion of an intrinsic pairing mechanism in the Hamiltonian of the nanowire (i.e., we put superconductivity into the nanowire by hand). Within this model of the proximity effect, we solve for the conductance analytically in three different limits: (1) in the absence of an external magnetic field, (2) when the strength of SOC is much larger than both the Zeeman splitting and the proximity-induced gap, and (3) when the Zeeman splitting is larger than both SOC and the gap. We show that the zero-bias conductance at zero temperature is fixed to $2e^2/h$ in the topological phase, and we also extend our calculation to finite temperature, calculating the conductance as a function of both energy and external field. 

In the second model of the proximity effect, superconductivity is induced in the nanowire through a tunnel coupling with a bulk conventional superconductor. Such a model allows for electrons to spend a fraction of their time in the underlying superconductor. We show that this model of the proximity effect has several interesting features. First, as pointed out also in Refs.~\onlinecite{Stanescu:2011,Potter:2011,Stanescu:2017}, the topological phase transition is determined by the strength of the tunnel coupling between the superconductor and nanowire rather than by the proximity-induced gap; therefore, very high magnetic fields are required to reach the topological phase if tunneling is made too strong. Second, the continuity equation is not obeyed in the nanowire if the excitation energy exceeds the gap of the superconductor; i.e., particles are lost from the nanowire to the substrate. Third, the conductance exhibits two distinct peaks as a function of energy; the first peak is located at the edge of the proximity-induced gap, while the second peak is located at the edge of the bulk gap of the superconductor. Finally, while the zero-bias conductance is fixed to $2e^2/h$ in the topological phase at zero temperature, we show that finite temperature can very drastically reduce the zero-bias conductance. Calculating the conductance as a function of energy and external field at finite temperature, we find that we can reproduce many of the qualitative features observed in Ref.~\onlinecite{Zhang:2016}.

The remainder of the paper is organized as follows. In Sec.~\ref{Sec2}, we calculate the conductance using a model that assumes an intrinsic pairing term in the Hamiltonian of the nanowire. We exactly solve for the conductance in the absence of an external field in Sec.~\ref{Zerofield}. In Sec.~\ref{FieldAnalytical}, we analytically solve for the conductance in the presence of an external field in two different limits: strong SOC (Sec.~\ref{Sec2}\,\ref{StrongSOC}) and strong field (Sec.~\ref{Sec2}\,\ref{StrongZeeman}). A numerical calculation of the conductance at finite temperature is presented in Sec.~\ref{Numerics}. In Sec.~\ref{Sec3}, we calculate the conductance using a model that accounts for tunneling between the superconducting substrate and the nanowire. We discuss an effective Hamiltonian describing proximity-induced superconductivity in the wire in Sec.~\ref{EffectiveH}. To illustrate the main qualitative features of this model, we analytically solve for the conductance in the absence of an external field in Sec.~\ref{Sec3}~\ref{ZeroFieldTunn}. Extension of the calculation to finite fields and finite temperatures is discussed in Sec.~\ref{Sec3}~\ref{FieldTunn}. Our conclusions are given in Sec.~\ref{Sec4}.

\section{Intrinsic pairing model} \label{Sec2}
\begin{figure}[t!]
\centering
\includegraphics[width=\linewidth]{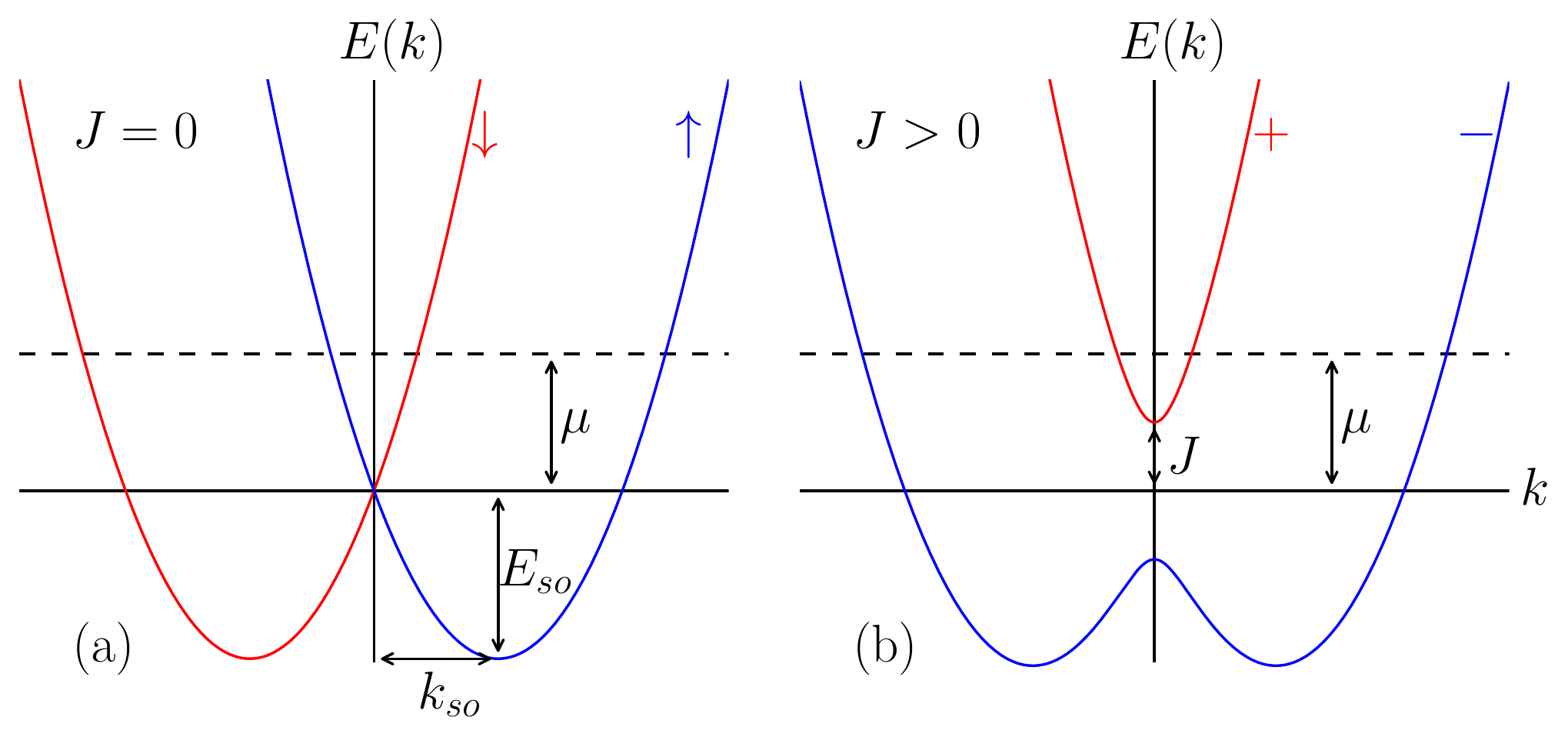}
\caption{\label{NormalSpectra}(a) 
Energy spectrum of a one-dimensional nanowire with Rashba spin-orbit coupling. The Fermi energy $\mu$ is measured from the degeneracy point at $k=0$. The spin-orbit coupling strength is parameterized by the spin-orbit energy $E_{so}=m\alpha^2/2$ and the spin-orbit wave vector $k_{so}=m\alpha$. (b) Energy spectrum of a Rashba nanowire in the presence of an external magnetic field applied parallel to the wire. Magnetic field opens a gap of size $2J$ at $k=0$.}
\end{figure}

We begin by discussing the intrinsic pairing model. The geometry we consider is displayed in Fig.~\ref{geometry}; an infinitely long 1D nanowire is placed on top of a junction between an insulator ($x<0$) and a superconductor ($x>0$), with an external magnetic field applied along the axis of the wire. The Hamiltonian of the system is taken to be
\begin{equation} \label{Hamiltonian}
H=H_{NW}+H_B+H_\Delta.
\end{equation}
The bare wire with Rashba-type \cite{Bychkov:1984} SOC is described by
\begin{equation} \label{HNW}
H_{NW}=\int dx\,\Psi^\dagger(x)\left(H_0+i\alpha\hat\sigma_z\partial_x\right)\Psi(x),
\end{equation}
where $H_0=-\partial_x^2/2m-\mu+U\delta(x)$, $\Psi(x)=[\psi_\uparrow(x),\psi_\downarrow(x)]^T$ is a spinor of second-quantized fermion operators describing states in the nanowire, and $\hat\sigma_i$ is a Pauli matrix acting in spin space. The effective mass $m$, Fermi energy $\mu$, and SOC constant $\alpha$ are taken to be constant throughout the length of the wire. We also allow for a delta-shaped barrier potential separating the superconducting and normal segments of the wire. The spin quantization axis is chosen along the direction of the effective Rashba field. In the bulk of the normal segment, the Rashba Hamiltonian yields an energy spectrum
\begin{equation} \label{RashbaSpectrum}
E_{\uparrow(\downarrow)}(k)=\xi_k\mp\alpha k,
\end{equation}
where $\xi_k=k^2/2m-\mu$. The spectrum of Eq.~(\ref{RashbaSpectrum}) is shown in Fig.~\ref{NormalSpectra}(a). SOC lifts the spin degeneracy of the wire at all momenta except $k=0$; the degeneracy at this point is preserved by time-reversal symmetry, and the Fermi energy $\mu$ is measured from this point. It is also convenient to parameterize the Rashba spectrum by the spin-orbit energy $E_{so}=m\alpha^2/2$ and the spin-orbit wave vector $k_{so}=m\alpha$, both of which are labeled in Fig.~\ref{NormalSpectra}(a); we will make use of this parameterization in what follows.

An external magnetic field parallel to the wire induces a term in the Hamiltonian given by
\begin{equation} \label{HB}
H_B=-{J}\int dx\,\Psi^\dagger(x)\hat\sigma_x\Psi(x),
\end{equation}
where ${J}=g\mu_BB_\text{ext}/2>0$ is the Zeeman energy in a field $B_\text{ext}$ ($g$ is the Land\'{e} g-factor and $\mu_B$ is the Bohr magneton). By breaking time-reversal symmetry, the external field lifts the degeneracy at $k=0$ and induces a gap of size $2J$ separating the two bands in the spectrum. Because this term contains $\hat\sigma_x$, spin is not a good quantum number in the presence of the field; we instead label the two bands by chirality indices $+$ and $-$:
\begin{equation}
E_\pm(k)=\xi_k\pm\sqrt{\alpha^2k^2+{J}}.
\end{equation}
The spectrum in the presence of the field is shown in Fig.~\ref{NormalSpectra}(b).

In this section, we model the proximity effect by an intrinsic pairing term induced in the nanowire. This term is described by a BCS-like Hamiltonian
\begin{equation}
H_\Delta=\int dx\,\Delta(x)\bigl[\psi_\downarrow(x)\psi_\uparrow(x)+H.c.\bigr],
\end{equation}
where we take $\Delta(x)=\Delta\theta(x)$; i.e., pairing is induced in only those parts of the nanowire contacting the underlying superconductor ($\Delta>0$ is assumed real and constant).

The Bogoliubov-de Gennes (BdG) equation describing the model is given by
\\
\begin{equation} \label{BdG}
\begin{aligned}
\biggl[H_0\hat\tau_z&+i\alpha\partial_x\hat\sigma_z-{J}\hat\tau_z\hat\sigma_x-\Delta(x)\hat\tau_y\hat\sigma_y\biggr]\psi(x)=E\psi(x),
\end{aligned}
\end{equation}
\\
where $\psi=[u_\uparrow,u_\downarrow,v_\uparrow,v_\downarrow]^T$ is the BdG spinor wave function describing states in the nanowire [$u(v)_\sigma$ is the wave function of an electron (hole) with spin $\sigma$], $\hat\tau_i$ are Pauli matrices acting in Nambu space, and $\hat\tau_i\hat\sigma_j=\hat\tau_i\otimes\hat\sigma_j$ is a Kronecker product of Pauli matrices. Following the BTK model, \cite{Blonder:1982} we look to solve Eq.~(\ref{BdG}) for the scattering wave function in both the normal ($N$) and superconducting ($S$) segments while imposing boundary conditions at the interface $x=0$. The boundary conditions can be obtained by directly integrating Eq.~(\ref{BdG}),
\begin{subequations} \label{BC2}
\begin{gather}
\psi_S(0)=\psi_N(0), \\
\partial_x\psi_S(0)-\partial_x\psi_N(0)=2mv_RZ\psi(0),
\end{gather}
\end{subequations}
where $v_R=\sqrt{\alpha^2+2\mu/m}$ is the Rashba velocity and $Z=U/v_R$ is a dimensionless barrier strength. We also identify the quasiparticle current,
\begin{equation} \label{QPcurrent}
\begin{aligned}
j(x)&=\frac{1}{m}\sum_\sigma\biggl\{\text{Im}\bigl[u_\sigma^*(x)\partial_xu_\sigma(x)-v^*_\sigma(x)\partial_xv_\sigma(x)\bigr] \\
	&-m\alpha\sigma\bigl[|u_\sigma(x)|^2+|v_\sigma(x)|^2\bigr]\biggr\},
\end{aligned}
\end{equation}
which is a conserved quantity of the BdG Hamiltonian.

We will now solve for the conductance within this model in several different limits. In Sec.~\ref{Zerofield} we solve for the conductance exactly in the absence of the external field. Using this solution, in Sec.~\ref{Sec2}~\ref{StrongSOC} we treat the external field perturbatively assuming that SOC is much larger than both the Zeeman splitting and the superconducting gap. This allows us to study the topological phase transition analytically. We are also able to treat the problem analytically deep in the topological phase, where the Zeeman splitting is very large; this is discussed in Sec.~\ref{Sec2}~\ref{StrongZeeman}. In Sec.~\ref{Numerics}, we relax all constraints on the parameters of the model. In doing so, we can no longer treat the problem analytically, so we resort to a numerical solution that we also extend to finite temperature. 

\begin{figure}[t!]
\centering
\includegraphics[width=\linewidth]{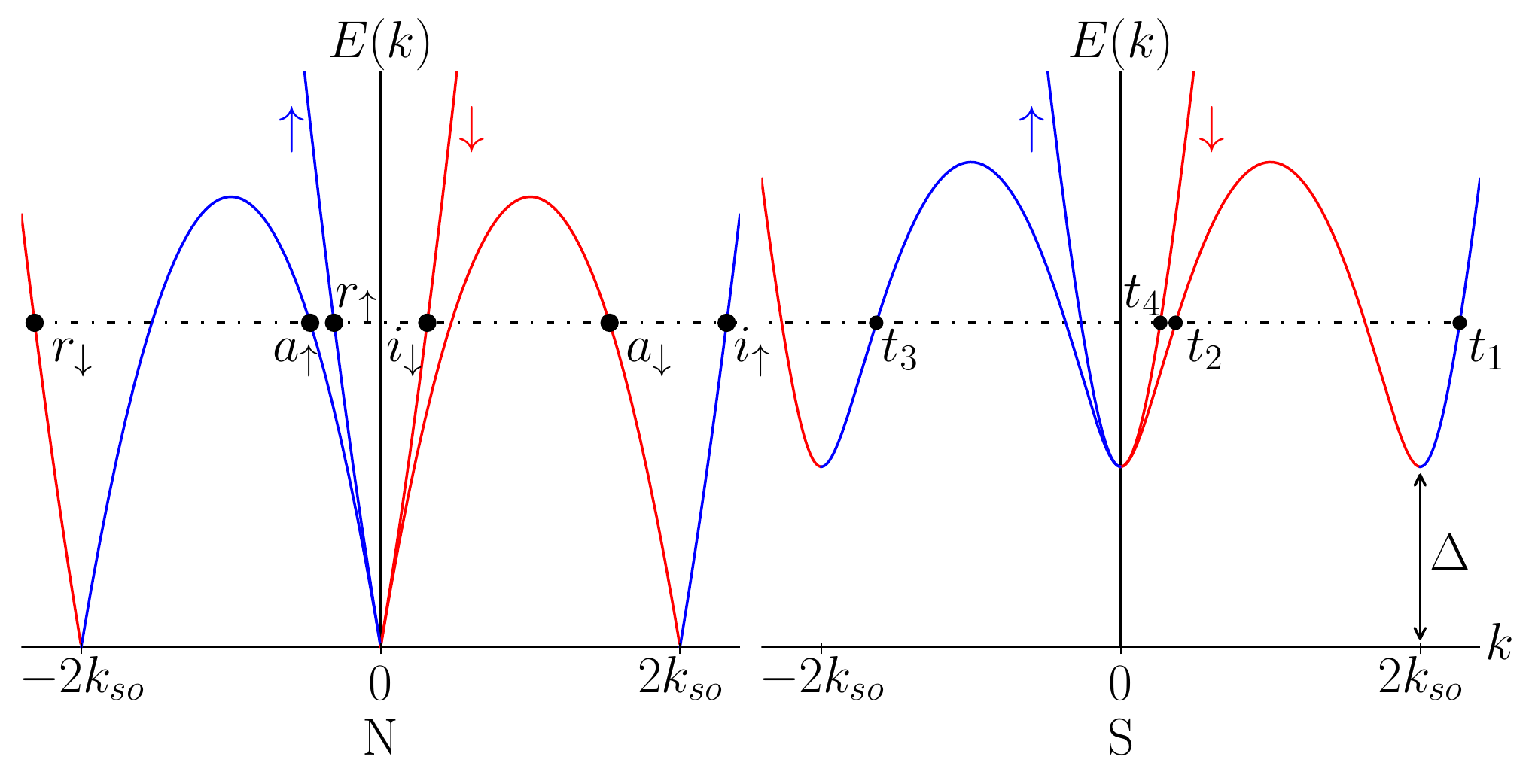}
\caption{\label{SNexcitation}Bogoliubov-de Gennes excitation spectra of normal (left) and superconducting (right) segments of one-dimensional nanowire with Rashba spin-orbit coupling in the absence of a magnetic field. Colors denote different spin states. An incident electron from the normal segment of the wire ($i_\sigma$) can be Andreev reflected ($a_\sigma$), normally reflected ($r_\sigma$), or transmitted to the superconducting segment ($t_i$). Spectra shown here taking $\mu=0$.}
\end{figure}

\subsection{Zero-Field Limit $B_\text{ext}=0$} \label{Zerofield}

First, we look to solve Eq.~(\ref{BdG}) when ${J}=0$. In this case, spin remains a good quantum number and the BdG equation yields eight eigenstates that can be characterized by spin, direction of propagation, and electron/hole character. In the normal segment, we find the momenta of these eight states by solving Eq.~(\ref{RashbaSpectrum}) for $k(E)$:
\begin{subequations} \label{momenta}
\begin{align}
k_{e\uparrow}^{R(L)}&=\pm\sqrt{2m(\mu+E_{so}+E)}+k_{so}, \\
k_{e\downarrow}^{R(L)}&=\pm\sqrt{2m(\mu+E_{so}+E)}-k_{so}, \\
k_{h\uparrow}^{R(L)}&=\mp\sqrt{2m(\mu+E_{so}-E)}-k_{so}, \\
k_{h\downarrow}^{R(L)}&=\mp\sqrt{2m(\mu+E_{so}-E)}+k_{so}.
\end{align}
\end{subequations}
The momenta of the eigenstates in the superconducting segment, $p_{e(h)\uparrow(\downarrow)}^{R(L)}$, are found from Eq.~(\ref{momenta}) by simply replacing $E\to\sqrt{E^2-\Delta^2}$. The BdG excitation spectra for both the normal and superconducting segments are shown in Fig.~\ref{SNexcitation}.

The scattering wave function in the normal segment is then found to be
\begin{equation}
\begin{aligned}
\psi_{N\sigma}(x)&=\psi_{i\sigma}(x)+a_{\uparrow\sigma}\left(\begin{array}{c} 0 \\ 0 \\ 1 \\ 0 \end{array}\right)e^{ik_{h\uparrow}^Lx}+a_{\downarrow\sigma}\left(\begin{array}{c} 0 \\ 0 \\ 0 \\ 1 \end{array}\right)e^{ik_{h\downarrow}^Lx} \\
	&+r_{\uparrow\sigma}\left(\begin{array}{c} 1 \\ 0 \\ 0 \\ 0 \end{array}\right)e^{ik_{e\uparrow}^Lx} + r_{\downarrow\sigma}\left(\begin{array}{c} 0 \\ 1 \\ 0 \\ 0 \end{array}\right)e^{ik_{e\downarrow}^Lx},
\end{aligned}
\end{equation}
where $\sigma$ refers to the spin of the incident particle and $a_{\sigma'\sigma}$ ($r_{\sigma'\sigma}$) denotes the Andreev (normal) reflection amplitude from an electron of spin $\sigma$ to a hole (electron) of spin $\sigma'$. There are two possibilities for the wave function of the incident particle: $\psi_{i\uparrow}(x)=(1,0,0,0)^Te^{ik_{e\uparrow}^Rx}$ or $\psi_{i\downarrow}(x)=(0,1,0,0)^Te^{ik_{e\downarrow}^Rx}$. The scattering wave function in the superconducting segment is given by
\begin{equation} \label{scatteringSRashba}
\begin{aligned}
\psi_{S\sigma}(x)&=t_{1\sigma}\left(\begin{array}{c} u_\Delta \\ 0 \\ 0 \\ v_\Delta \end{array}\right)e^{ip_{e\uparrow}^Rx}+t_{2\sigma}\left(\begin{array}{c} 0 \\ u_\Delta \\ -v_\Delta \\ 0 \end{array}\right)e^{ip_{e\downarrow}^Rx} \\
	&+t_{3\sigma}\left(\begin{array}{c} 0 \\ -v_\Delta \\ u_\Delta \\ 0 \end{array}\right)e^{ip_{h\uparrow}^Rx}+t_{4\sigma}\left(\begin{array}{c} v_\Delta \\ 0 \\ 0 \\ u_\Delta \end{array}\right)e^{ip_{h\downarrow}^Rx},
\end{aligned}
\end{equation}
where we define a generalization of the usual BCS coherence factors
\begin{subequations} \label{coherence}
\begin{align}
u_\lambda&=\frac{\sgn{\lambda}}{\sqrt{2}}\left(1+\frac{\sqrt{E^2-\lambda^2}}{E}\right)^{1/2}, \\
v_\lambda&=\frac{1}{\sqrt{2}}\left(1-\frac{\sqrt{E^2-\lambda^2}}{E}\right)^{1/2}. 
\end{align}
\end{subequations}
In Eqs.~(\ref{coherence}), $\lambda$ is an energy scale (specified by the subscript of $u$ and $v$) that will take several different values throughout the remainder of the paper. For example, the coherence factors in Eq.~(\ref{scatteringSRashba}) take $\lambda=\Delta$ [with $\sgn\lambda=1$].

Suppose first that a spin-up electron is incident from the normal segment on the superconducting interface. If we assume that $(\mu+E_{so})\gg\Delta$ (i.e., the Fermi level is not too close to the bottom of the Rashba band), then we can make a semiclassical approximation whereby $k_{e\uparrow}^R=p_{e\uparrow}^R=k_{h\downarrow}^L=m(v_R+\alpha)$ and $k_{h\uparrow}^L=p_{e\downarrow}^R=m(v_R-\alpha)$. Solutions to Eqs.~(\ref{BC2}) are then given by
\begin{subequations} \label{scatteringamps}
\begin{gather}
a_{\downarrow\uparrow}=\frac{u_\Delta v_\Delta}{u_\Delta^2+(u_\Delta^2-v_\Delta^2)Z^2}, \\
r_{\uparrow\uparrow}=-\frac{(u_\Delta^2-v_\Delta^2)Z(i+Z)}{u_\Delta^2+(u_\Delta^2-v_\Delta^2)Z^2}, \\
t_{1\uparrow}=\frac{u_\Delta(1-iZ)}{u_\Delta^2+(u_\Delta^2-v_\Delta^2)Z^2}, \\
t_{4\uparrow}=\frac{iv_\Delta Z}{u_\Delta^2+(u_\Delta^2-v_\Delta^2)Z^2},
\end{gather}
\end{subequations}
with the remaining scattering amplitudes equal to 0 (i.e., the two spin channels scatter independently of one another). We note that the scattering amplitudes given in Eqs.~(\ref{scatteringamps}) are precisely the same as those found in a superconductor/normal junction without SOC \cite{Blonder:1982} (SOC enters the solution only through the renormalization of the dimensionless barrier strength $Z=U/v_R$). If we instead have an incident spin-down electron, the scattering amplitudes are found from Eqs.~(\ref{scatteringamps}) by replacing $a_{\downarrow\uparrow}\to-a_{\uparrow\downarrow}$, $r_{\uparrow\uparrow}\to r_{\downarrow\downarrow}$, $t_{1\uparrow}\to t_{2\downarrow}$, and $t_{4\uparrow}\to-t_{3\downarrow}$. \cite{Reeg:2015}

The conductance is calculated from the various quasiparticle currents [defined in Eq.~(\ref{QPcurrent})] carried by each of the scattering states. In the semiclassical limit, both incident spin states carry the same current, $j_{\sigma}^i=v_R$. The currents carried by Andreev and normally reflected states are $j_{\sigma'\sigma}^a=-v_R|a_{\sigma'\sigma}|^2$ and $j_{\sigma'\sigma}^r=-v_R|r_{\sigma'\sigma}|^2$, respectively. Therefore, the conductance takes the simple form
\begin{equation} \label{Gnofield}
G(E)=\frac{e^2}{h}\sum_{\sigma,\sigma'}\left[2+|a_{\sigma'\sigma}|^2-|r_{\sigma'\sigma}|^2\right],
\end{equation}
where $e$ is the electron charge and $h$ is Planck's constant. Again, SOC modifies the conductance of an SN junction \cite{Blonder:1982} only through the renormalization of $Z$. At zero energy, the conductance is given by
\begin{equation} \label{G0nofield}
G(0)=\frac{2e^2}{h}\frac{2}{(1+2Z^2)^2}.
\end{equation}

\subsection{Analytic Solutions with External Field} \label{FieldAnalytical}
If we introduce the external magnetic field, the excitation spectrum in the bulk of the superconducting segment of the wire takes the form \cite{Oreg:2010,Lutchyn:2010}
\begin{equation} \label{topspectrum}
\begin{aligned}
E_\pm^2(k)&={J}^2+\Delta^2+\xi_k^2+(\alpha k)^2 \\
	&\pm2\sqrt{{J}^2\Delta^2+{J}^2\xi_k^2+(\alpha k)^2\xi_k^2}.
\end{aligned}
\end{equation}
At $k=0$, the lower branch of the spectrum $E_-$ has an excitation gap of $|\sqrt{\Delta^2+\mu^2}-{J}|$ which closes and reopens upon increasing the strength of the field. The critical field strength ${J}_c=\sqrt{\Delta^2+\mu^2}$, where the gap closes, marks a topological phase transition. For fields ${J}>{J}_c$, the superconducting segment of the wire is in the topological phase and supports a Majorana fermion mode localized to its boundary (in our geometry, this corresponds to the SN interface). Studying transport within this model requires a solution for the momenta of the scattering eigenstates; however, solving Eq.~(\ref{topspectrum}) analytically for $k(E)$ assuming an arbitrary set of parameters gives a very complicated result. In order to proceed analytically, we will treat the special cases of strong spin-orbit coupling and strong magnetic field. For simplicity, we also set $\mu=0$ throughout our analytical calculations. 

We note that the following analytical calculations are very similar to those presented in Ref.~\onlinecite{Setiawan:2015}. However, the model considered in Ref.~\onlinecite{Setiawan:2015} is not equivalent to ours, as the normal segment is not subjected to spin-orbit coupling or an external magnetic field (i.e., the normal segment consists of two degenerate spin channels).

\subsubsection{Strong Spin-Orbit Coupling ($E_{so}\gg {J},\Delta$)} \label{StrongSOC}
\begin{figure}[t!]
\centering
\includegraphics[width=\linewidth]{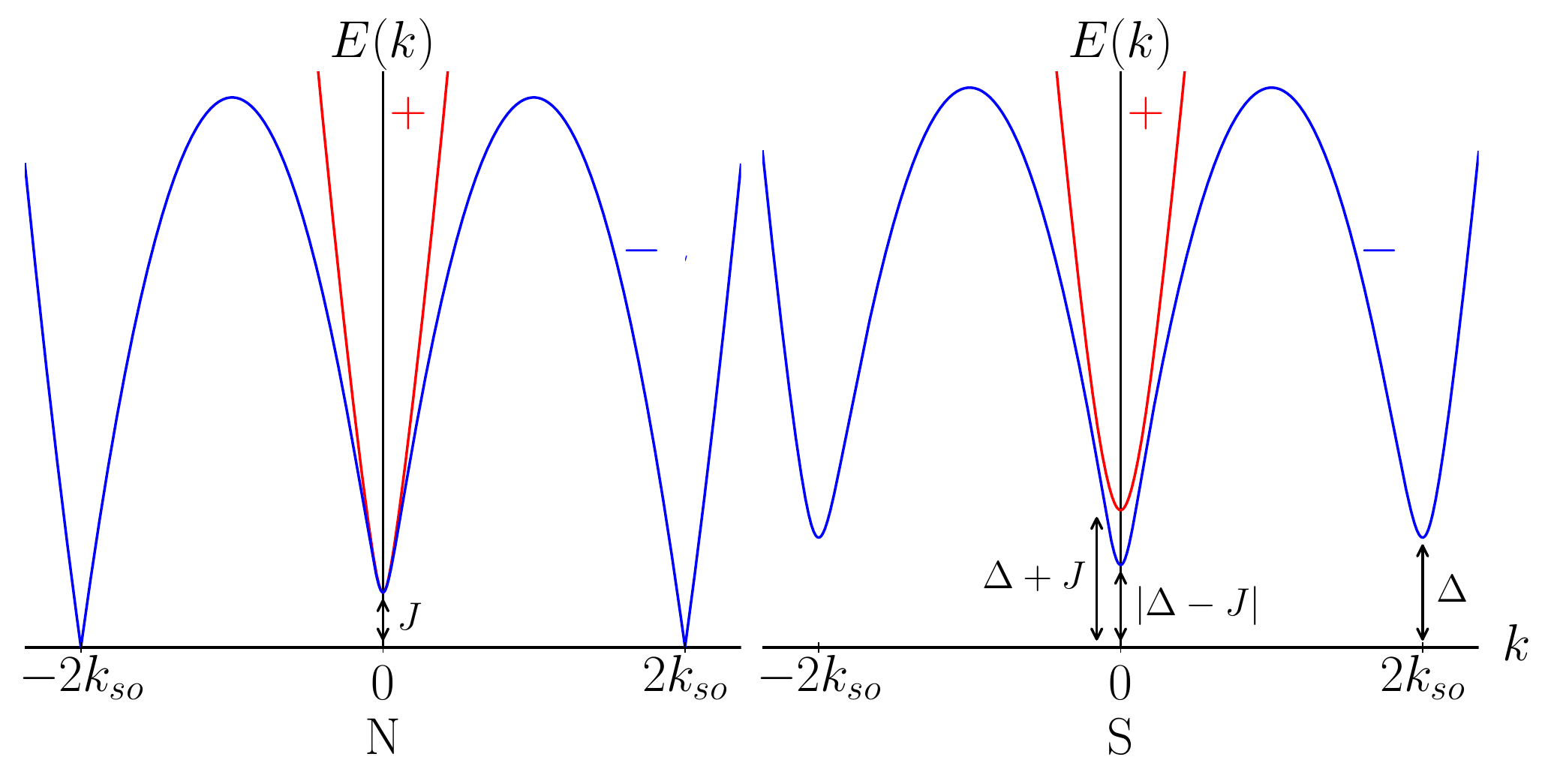}
\caption{\label{SNexcitation2}Bogoliubov-de Gennes excitation spectra of normal (left) and superconducting (right) segments of nanowire in the presence of an external field in the intrinsic pairing model. Chemical potential is fixed at $\mu=0$ and $E_{so}\gg {J},\Delta$. The spectra are shown here for $J/\Delta=0.25$.}
\end{figure}

In the normal segment of the wire ($\Delta=0$), it is possible to solve Eq.~(\ref{topspectrum}) and obtain a relatively simple expression for $k(E)$ for an arbitrary strength of SOC. Expanding this solution in the limit ${J}\ll E_{so}$ gives a total of eight possible momenta that take the form
\begin{subequations}
\begin{gather}
k_{e-}=k_{h-}=2k_{so}, \\
k_{e+}=k_{h+}=\sqrt{E^2-{J}^2}/\alpha,
\end{gather}
\end{subequations}
with momenta of opposite sign also being eigenstates. To lowest order, states near the Fermi momentum of the lower subband are unaffected by the presence of the field, while states near $k=0$ are strongly affected (see Fig.~\ref{SNexcitation2}). The scattering wave function is given by
\begin{equation} \label{scatteringN}
\begin{aligned}
\psi_{N}(x)&=\psi_{i\pm}(x)+a_{-\pm}\left(\begin{array}{c} 0 \\ 0 \\ 0 \\ 1 \end{array}\right)e^{ik_{h-}x}+r_{-\pm}\left(\begin{array}{c} 0 \\ 1 \\ 0 \\ 0 \end{array}\right)e^{-ik_{e-}x} \\
	&+a_{+\pm}\left(\begin{array}{c} 0 \\ 0 \\ u_{J} \\ v_{{J}} \end{array}\right)e^{-ik_{h+}x}+r_{+\pm}\left(\begin{array}{c} u_{J} \\ -v_{J} \\ 0 \\ 0 \end{array}\right)e^{-ik_{e+}x},
\end{aligned}
\end{equation}
where, for example, $a_{+-}$ ($r_{+-}$) denotes the amplitude of Andreev (normal) reflection from an electron of momentum $k_{e-}$ to a hole (electron) of momentum $k_{h(e)+}$, and $u_J$ and $v_{J}$ are generalized coherence factors as defined in Eqs.~(\ref{coherence}) taking $\lambda=J$ [with $\sgn \lambda=1$]. The two possible incident states are given by $\psi_{i-}=(1,0,0,0)^Te^{ik_{e-}x}$ and $\psi_{i+}=(v_{J},-u_{J},0,0)^Te^{ik_{e+}x}$. However, for $E<{J}$, the momentum $k_{e+}$ is imaginary and there is only one possible conducting channel in the normal segment.

In the superconducting segment, we solve Eq.~(\ref{topspectrum}) for $k(E)$ in the strong spin-orbit limit to give eight possible momenta. The four allowed momenta that enter the scattering wave function are given by
\begin{subequations}
\begin{gather}
p_{e-}=2k_{so}+\sqrt{E^2-\Delta^2}/\alpha, \\
-p_{h-}=-2k_{so}+\sqrt{E^2-\Delta^2}/\alpha, \\
p_{e+}=\sqrt{E^2-(\Delta+ {J})^2}/\alpha, \\
p_{h+}=\sqrt{E^2-(\Delta-{J})^2}/\alpha.
\end{gather}
\end{subequations}
As shown in Fig.~\ref{SNexcitation2}, the spectrum in the bulk of the superconducting segment has three different excitation gaps. States near the Fermi momentum have a gap of $\Delta$, while electrons near $k=0$ have a gap of $\Delta+{J}$ and holes near $k=0$ have a gap of $|\Delta-{J}|$. The scattering wave function in the superconducting segment is given by
\begin{equation} \label{scatteringS}
\begin{aligned}
\psi_S(x)&=t_{e-}\left(\begin{array}{c} u_\Delta \\ 0 \\ 0 \\ v_\Delta \end{array}\right)e^{ip_{e-}x}+\frac{t_{e+}}{\sqrt{2}}\left(\begin{array}{c} v_{\Delta+{J}} \\ -u_{\Delta+{J}} \\ v_{\Delta+{J}} \\ u_{\Delta+{J}} \end{array}\right)e^{ip_{e+}x} \\
	&+t_{h-}\left(\begin{array}{c} 0 \\ -v_\Delta \\ u_\Delta \\ 0 \end{array}\right)e^{-ip_{h-}x}+\frac{t_{h+}}{\sqrt{2}}\left(\begin{array}{c} v_{\Delta-{J}} \\ u_{\Delta-{J}} \\ -v_{\Delta-{J}} \\ u_{\Delta-{J}} \end{array}\right)e^{ip_{h+}x}.
\end{aligned}
\end{equation}

First, let us calculate the conductance through the junction for $E<{J}$. In this case, we only need to consider a single incident scattering channel, corresponding to $\psi_{i-}$ in Eq.~(\ref{scatteringN}), and the conductance takes the form
\begin{equation} \label{Gless}
G(E<{J})=\frac{e^2}{h}\left[1+|a_{--}|^2-|r_{--}|^2\right].
\end{equation}
Boundary conditions [Eqs.~(\ref{BC2})] can be solved analytically to lowest order in $1/E_{so}$, but the resulting expressions for the scattering amplitudes are very cumbersome; we spell out both the boundary conditions and resulting scattering amplitudes in Appendix~\ref{BCapp} rather than here. 

\begin{figure}
\centering
\includegraphics[width=\linewidth]{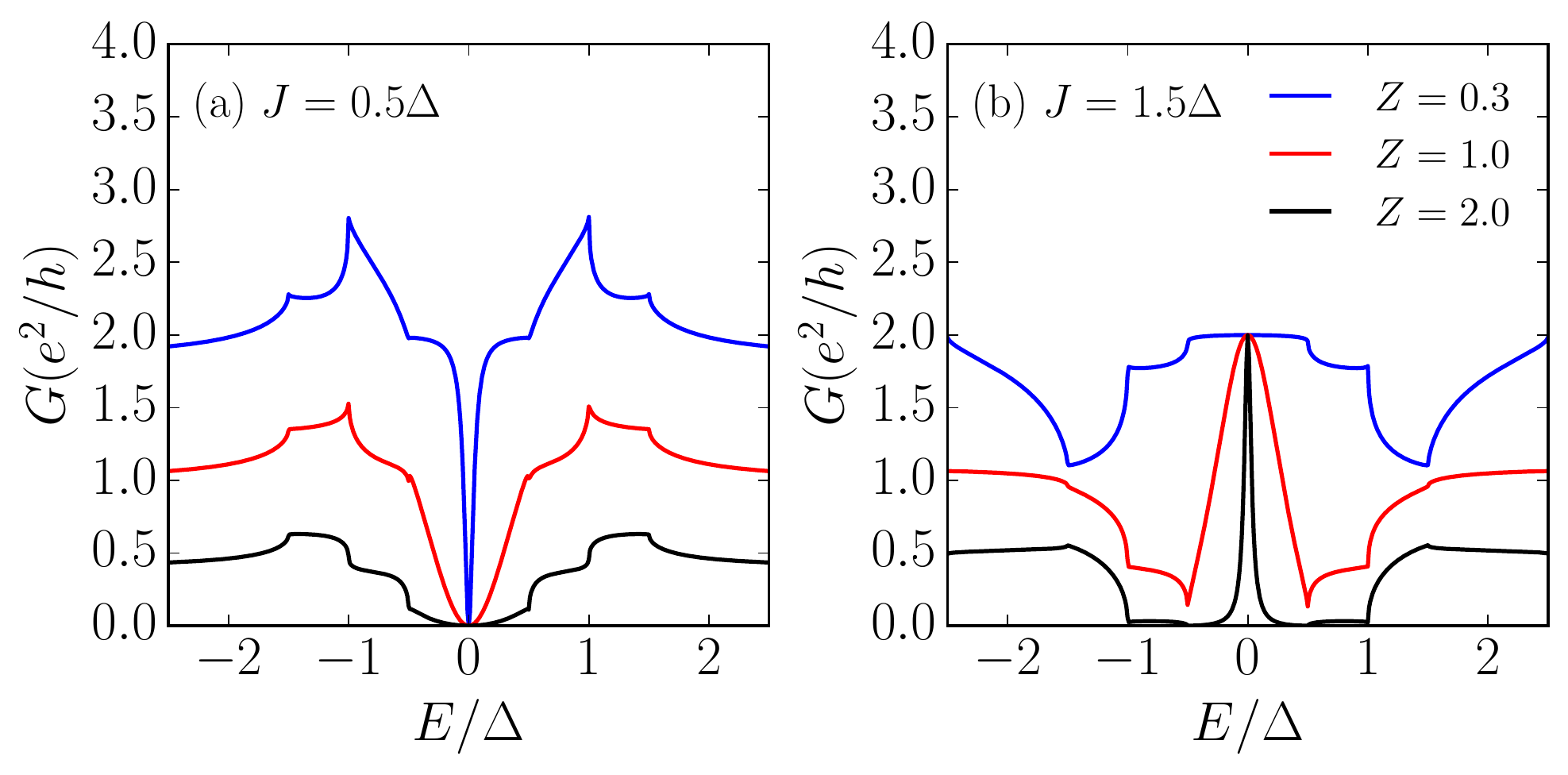}
\caption{\label{condT0}Conductance spectra in strong spin-orbit limit ($E_{so}\gg J,\Delta$), plotted for several values of $Z$ with fixed $J/\Delta$ at zero temperature. (a) Nontopological phase (${J}=0.5\Delta$). (b) Topological phase (${J}=1.5\Delta$). Chemical potential is fixed at $\mu=0$.}
\end{figure}

However, the conductance takes a particularly simple form at $E=0$:
\begin{equation} \label{G0field}
G(0)=\frac{e^2}{h}\left\{\begin{array}{rc}
	0, & 0<{J}<\Delta \\
	\displaystyle \frac{2}{1+Z^4(1+Z^2)^2}, & {J}=\Delta \\
	2, & {J}>\Delta \end{array}\right..
\end{equation}
The conductance takes the universal values 0 and $2e^2/h$ in the nontopological and topological phases, respectively, while taking a nonuniversal value (dependent on the barrier strength $Z$) at the phase transition ${J}=\Delta$. While Eq.~(\ref{G0field}) was obtained to lowest order in $1/E_{so}$, we now show that the universal values for the zero-bias conductance hold to all orders (i.e., away from the strong SOC limit). To do so, we adapt the scattering matrix theory of Ref.~\onlinecite{Law:2009} for a spinless normal/superconductor junction to the problem currently under consideration. Assuming that the Fermi level lies inside the field-induced gap at $k=0$ (this is always true for $\mu=0$ and $J>0$), the scattering problem can be recast in terms of a scattering matrix by
\begin{equation} \label{scatteringmatrix}
\left(\begin{array}{c} \psi_{e-}^L(E) \\ \psi_{h-}^L(E) \end{array}\right)=\left(\begin{array}{cc}
	r_{--}(E) & \bar a_{--}(E) \\ a_{--}(E) & \bar r_{--}(E) \end{array}\right)\left(\begin{array}{c} \psi_{e-}^R(E) \\ \psi_{h-}^R(E) \end{array}\right),
\end{equation}
where $\psi_{e-}^{R(L)}=[u_{\uparrow(\downarrow)},u_{\downarrow(\uparrow)}]^Te^{\pm ik_{e-}x}$ are the incident (reflected) electron states in the lower subband [the particular forms of $u_{\uparrow(\downarrow)}$ away from the strong SOC limit are unimportant and left unspecified; also note that states of opposite momentum are related by flipping the spin in Eq.~(\ref{BdG})]. Similarly, the incident (reflected) hole states are given by $\psi_{h-}^{R(L)}=[v_{\uparrow(\downarrow)},v_{\downarrow(\uparrow)}]^Te^{\mp ik_{h-}x}$. The reflection amplitudes $a_{--}$ and $r_{--}$ are the same as in Eq.~(\ref{scatteringN}), while their counterparts $\bar a_{--}$ and $\bar r_{--}$ indicate that the incident state is a hole rather than an electron. Taking the upper components of the spinors, Eq.~(\ref{scatteringmatrix}) can be expressed as
\begin{subequations}
\begin{align}
u_\downarrow(E)=r_{--}(E)u_\uparrow(E)+\bar a_{--}(E)v_\uparrow(E), \\
v_\downarrow(E)=a_{--}(E)u_\uparrow(E)+\bar r_{--}(E)v_\uparrow(E).
\end{align}
\end{subequations}
Particle-hole symmetry of the BdG equation [Eq.~(\ref{BdG})] dictates that the electron and hole wave functions are related through $u_\sigma(E)=v^*_\sigma(-E)$; this in turn imposes a constraint on the scattering matrix through the relations $r_{--}(E)=\bar r_{--}^*(-E)$ and $a_{--}(E)=\bar a_{--}^*(-E)$. Therefore, at $E=0$, unitarity of scattering matrix requires $a_{--}(0)=0$ or $r_{--}(0)=0$; i.e., either the incident state undergoes perfect normal or perfect Andreev reflection. Of course, this scattering matrix argument breaks down for $J=0$ (when states near $k=0$ in normal segment are available for scattering at $E=0$) and for $J=J_c$ (when the gap in the superconducting segment closes and transmission at $E=0$ is possible). It is only for these specific values of $J$ that the zero-bias conductance can take nonuniversal values [corresponding to Eq.~(\ref{G0nofield}) and the $J=\Delta$ case of Eq. (\ref{G0field}), respectively]. Additionally, we can extend the scattering matrix arguments to cover the case $\mu\neq0$; in this case, $G(0)=0$ for $|\mu|<J<J_c$ and $G(0)=2e^2/h$ for $J>J_c$.

\begin{figure}
\centering
\includegraphics[width=\linewidth]{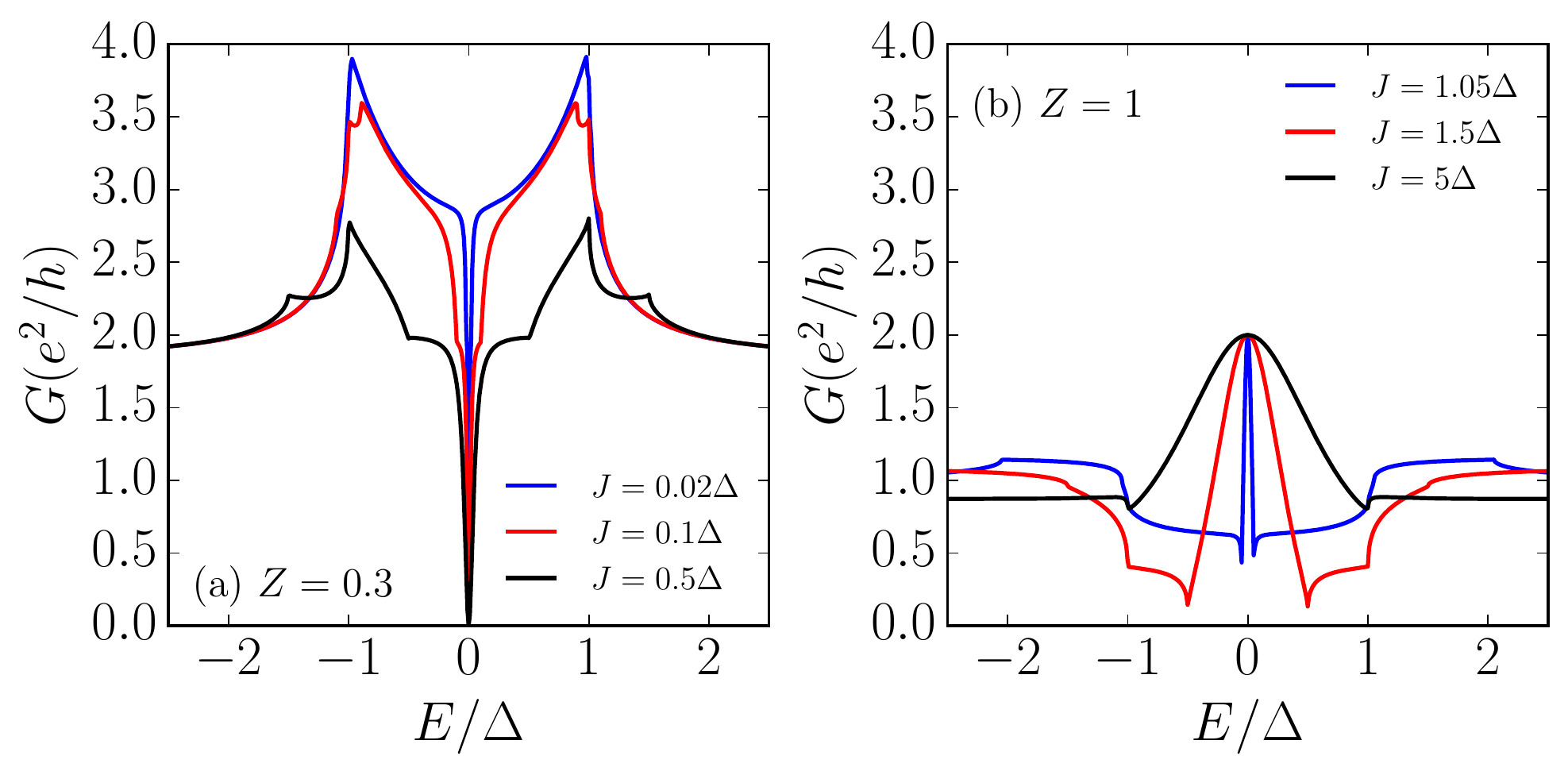}
\caption{\label{condT02}Conductance spectra in strong spin-orbit limit ($E_{so}\gg J,\Delta$), plotted for several values of $J/\Delta$ with fixed $Z$ at zero temperature. (a) Nontopological phase (${J}<\Delta$) with $Z=0.3$. (b) Topological phase (${J}>\Delta$) with $Z=1$. Chemical potential is fixed at $\mu=0$.}
\end{figure}

When the excitation energy exceeds the Zeeman splitting ($E>{J}$), there are two possible incident scattering channels. Therefore, to calculate the conductance we must also consider the incident state corresponding to $\psi_{i+}$ in Eq.~(\ref{scatteringN}). We spell out the explicit solutions to this scattering problem in Appendix~\ref{BCapp}. Accounting for both conducting channels, the conductance is given by
\begin{equation} \label{Gmore}
\begin{aligned}
G(E>{J})&=\frac{e^2}{h}\biggl[2+|a_{--}|^2+|a_{++}|^2-|r_{--}|^2 \\
	&-|r_{++}|^2+(u_{J}^2-v_{J}^2)(|a_{+-}|^2-|r_{+-}|^2) \\
	&+\frac{1}{u_{J}^2-v_{J}^2}(|a_{-+}|^2-|r_{-+}|^2)\biggr].
\end{aligned}
\end{equation}

The conductance is plotted as a function of energy in both the nontopological and topological regimes in Figs.~\ref{condT0} and \ref{condT02}. In the nontopological phase the conductance exhibits a dip at low energies, while the conductance exhibits a peak in the topological phase. In both instances, the width of the low-energy feature is controlled by both parameters of the problem ($Z$ and $J/
\Delta$, with SOC entering implicitly through the definition $Z=U/\alpha$). In Fig.~\ref{condT0}, we plot the conductance at fixed $J/\Delta$ for various values of $Z$. We find that in the nontopological phase, the width of the zero-bias dip is decreased with decreasing $Z$. In Fig.~\ref{condT02} we demonstrate the dependence of the low-energy feature on $J/\Delta$. In both phases, the conductance exhibits sharp features at the four energies corresponding to gap edges in the spectra of both the normal and superconducting segments; these energies correspond to $J$, $\Delta$, $|\Delta-J|$, and $\Delta+J$, as labeled in Fig.~\ref{SNexcitation2}.

\subsubsection{Strong Magnetic Field (${J}\gg E_{so},\Delta$)} \label{StrongZeeman}
\begin{figure}[t!]
\centering
\includegraphics[width=\linewidth]{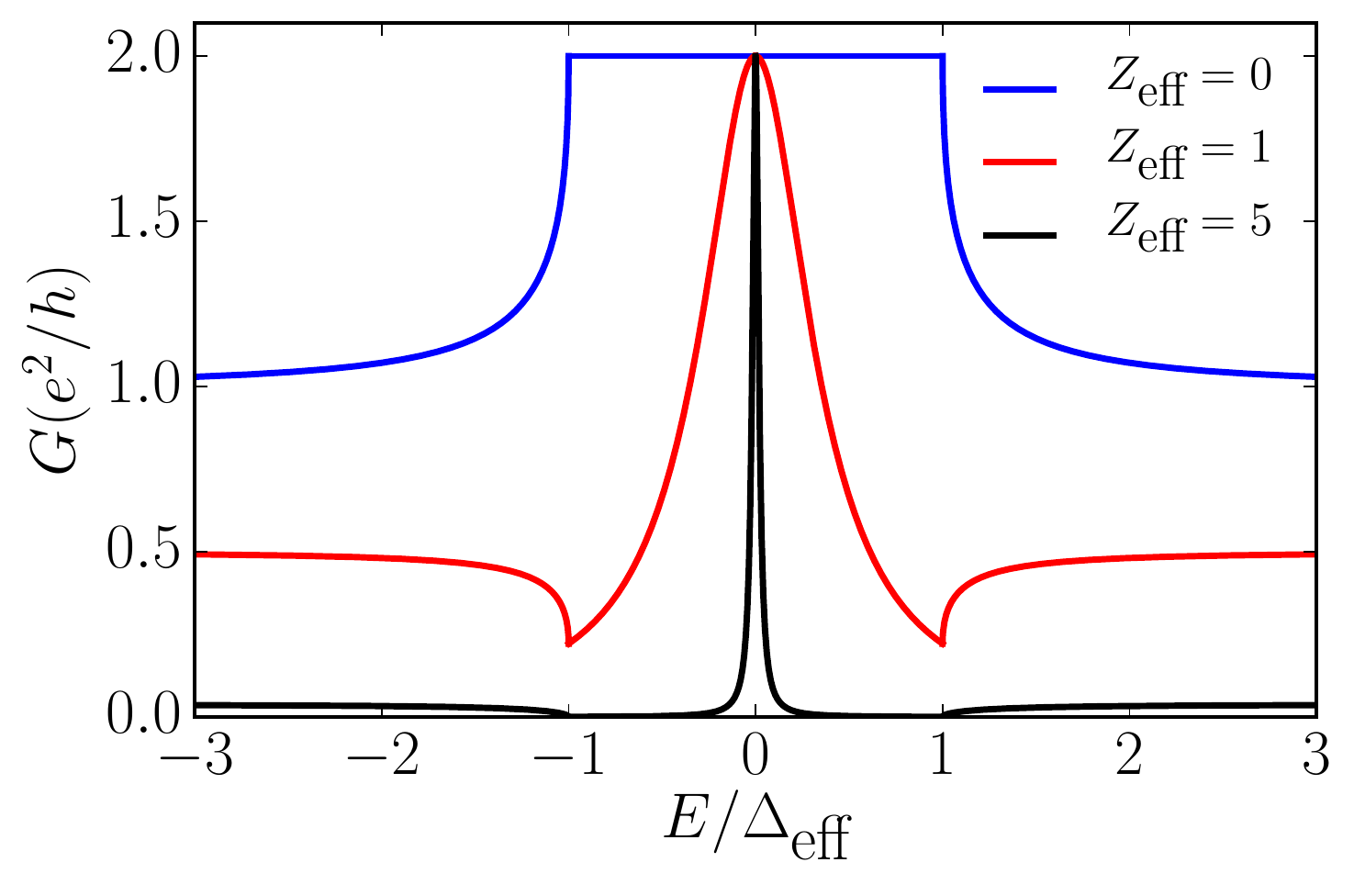}
\caption{\label{pwaveplot}Conductance deep in the topological phase ($J\gg E_{so},\Delta$), plotted for several values of barrier strength $Z_\text{eff}=Z\sqrt{E_{so}/J}$. The zero-bias conductance is fixed at $2e^2/h$ and the width of the zero-bias peak is controlled by $Z_\text{eff}$.}
\end{figure}

In the limit of a strong external field ${J}\gg E_{so},\Delta$, the two subbands are nearly spin-polarized and the upper subband plays no role in transport. Projecting out the upper subband, we can map the Hamiltonian (\ref{Hamiltonian}) directly onto the low-density limit of the Kitaev model for spinless $p$-wave superconductivity,\cite{Kitaev:2001}
\begin{equation} \label{Kitaev}
\begin{aligned}
H&=\int dx\biggl\{\psi^\dagger(x)\left(-\frac{\partial_x^2}{2m}-\mu_\text{eff}\right)\psi(x) \\
	&+\bigl[\Delta_\text{eff}(x)\psi^\dagger(x)(-i\partial_x/k_{F,\text{eff}})\psi^\dagger(x)+H.c.\bigr]\biggr\},
\end{aligned}
\end{equation}
where $\mu_\text{eff}={J}$, $k_{F,\text{eff}}=\sqrt{2mJ}$, and $\Delta_\text{eff}(x)=i\Delta\sqrt{E_{so}/J}\theta(x)\equiv i\Delta_\text{eff}\theta(x)$. \cite{Alicea:2011} While the conductance through a junction between a spinless normal metal and spinless $p$-wave superconductor has been studied in several other works, \cite{Yan:2014,Setiawan:2015,Thakurathi:2015,Zazunov:2016} most of these studies focus on the conductance near the topological phase transition $\mu_\text{eff}=0$. However, in the strong-field limit we are deep in the topological phase and the limit $\mu_\text{eff}\gg\Delta_\text{eff}$ is more relevant. In this limit, it is possible to apply a semiclassical approximation (similarly to what can be done for a conventional normal metal/$s$-wave superconductor junction \cite{Blonder:1982}), and the full simplified semiclassical calculation is presented in Appendix~\ref{spinlessapp}. The conductance is given by
\begin{subequations} \label{pwavecond}
\begin{align}
G(E<\Delta_\text{eff})&=\frac{2e^2}{h}\frac{W^2}{W^2+4E^2}, \\
G(E>\Delta_\text{eff})&=\frac{e^2}{h}\frac{2E[E+\Omega_\text{eff}(1+2Z_\text{eff}^2)]}{[E(1+2Z_\text{eff}^2)+\Omega_\text{eff}]^2},
\end{align}
\end{subequations}
where $Z_\text{eff}=Z\sqrt{E_{so}/J}$ and $\Omega_\text{eff}=\sqrt{E^2-\Delta_\text{eff}^2}$, and is plotted for several values of $Z_\text{eff}$ in Fig.~\ref{pwaveplot}. The subgap conductance takes a Lorentzian form with amplitude $2e^2/h$ and width $W=\Delta_\text{eff}/\sqrt{Z_\text{eff}^2(1+Z_\text{eff}^2)}$.

\subsection{Finite Temperature} \label{Numerics}
We now incorporate the effects of finite temperature in our calculation in two ways. First, the pairing potential is assumed to follow a BCS-like temperature dependence, which is implemented using the interpolation function
\begin{equation} \label{DeltaT}
\Delta(T)=\Delta(0)\tanh\left(1.74\sqrt{T_c/T-1}\right),
\end{equation}
where $T_c$ is the critical temperature such that $\Delta(T_c)=0$. Second, finite temperature broadens the Fermi function so that the finite-temperature conductance is given by
\begin{equation} \label{GT}
G(E)=\int d\varepsilon\,G_0(\varepsilon)\left(-\frac{\partial f(\varepsilon,E)}{\partial\varepsilon}\right),
\end{equation}
where $G_0(\varepsilon)$ is the zero-temperature conductance and $f(\varepsilon,E)=\{1+\exp[(\varepsilon-E)/T]\}^{-1}$ is the Fermi function. Notably, we do not incorporate inelastic scattering or dephasing processes induced by finite temperature into our model.

\begin{figure}[t!]
\centering
\includegraphics[width=\linewidth]{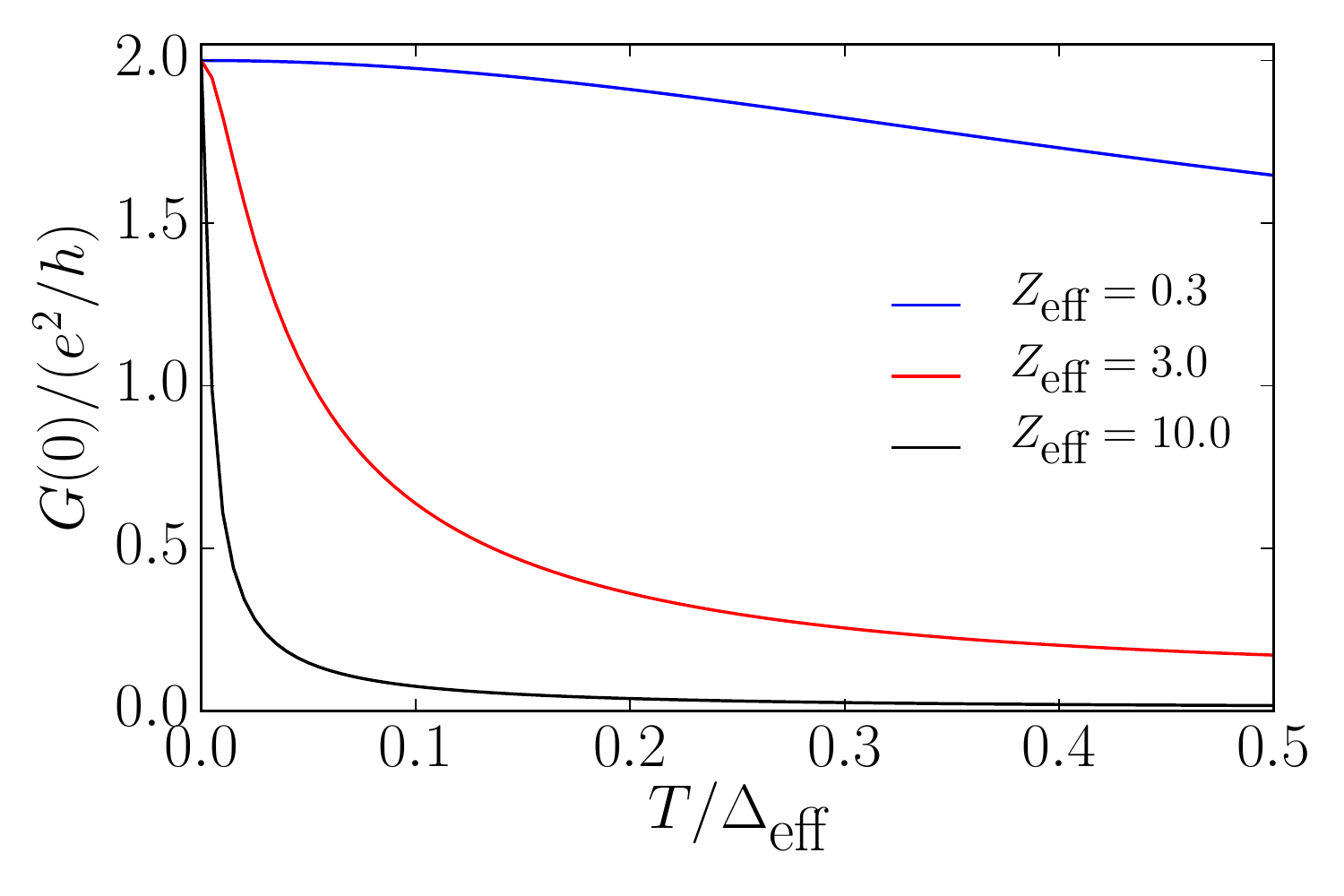}
\caption{\label{G0vsT}Zero-bias conductance $G(0)$ as a function of temperature deep in the topological phase ($J\gg E_{so},\Delta$), plotted for several values of $Z_\text{eff}=Z\sqrt{E_{so}/J}$. At $T=0$ the zero-bias conductance is $2e^2/h$, while finite temperature more significantly alters the conductance as the effective barrier strength is increased.}
\end{figure}

Deep in the topological phase (see Sec.~\ref{Sec2}\,\ref{StrongZeeman}), we investigate the dependence of the zero-bias conductance on the temperature $T$.  If we assume that we are in the tunneling limit $Z_\text{eff}\gg1$ and at low temperatures $T\ll\Delta_\text{eff}\ll\Delta$, the zero-bias conductance is given by
\begin{equation} \label{zerobiasfiniteT}
G(0)=\frac{2e^2}{h}\int d\varepsilon\left(\frac{1}{\pi}\frac{W}{W^2+4\varepsilon^2}\right)\frac{\pi W}{4T\cosh^2(\varepsilon/2T)},
\end{equation}
where we approximate $W=\Delta_\text{eff}/Z_\text{eff}^2$ for $Z_\text{eff}\gg1$. The $T$-dependence of the conductance is the same as for tunneling through a resonant level. When $T\ll W$, the derivative of the Fermi function can be replaced by a $\delta$-function and the  zero-bias conductance is given by $G(0)=2e^2/h$. When $W\ll T\ll\Delta_\text{eff}$, the quantity in parentheses in Eq. (\ref{zerobiasfiniteT}) can be replaced by $\delta(\varepsilon)/2$ and the conductance
 is given by
\begin{equation} \label{G0finiteT}
G(0)=\frac{e^2}{h}\frac{\pi W}{4T}\ll\frac{2e^2}{h}.
\end{equation}
The zero-bias conductance in this limit falls as $1/T$ as the temperature is increased (this result was also seen numerically in Ref.~\onlinecite{Chevallier:2016}). Therefore, the temperatures at which one achieves the expected universal zero-bias conductance $2e^2/h$ is determined by both the strength of the tunnel barrier and the ratio $E_{so}/J$. This is demonstrated in Fig.~\ref{G0vsT}, where we plot the zero-bias conductance as a function of $T$, calculated numerically by substituting Eq.~(\ref{pwavecond}) into Eq.~(\ref{GT}). As demonstrated analytically, temperature has a much more severe effect on the zero-bias conductance in the tunneling limit.

\begin{figure*}[t!]
\centering
\includegraphics[width=0.9\linewidth]{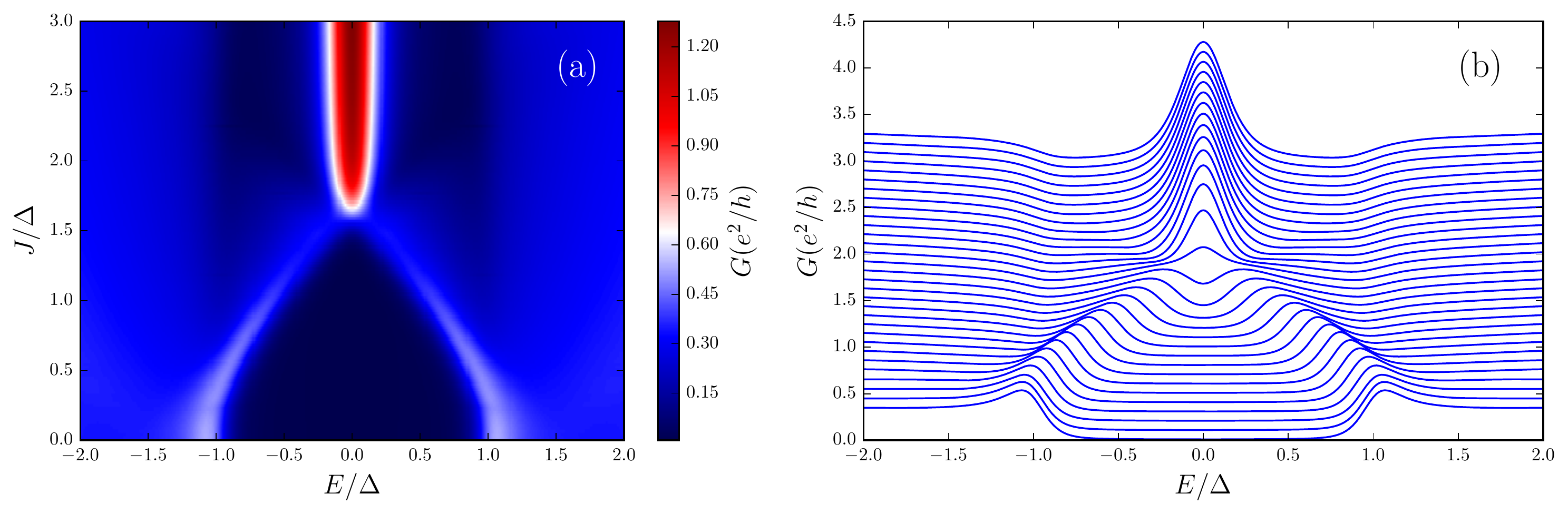}
\caption{\label{finiteT}(a) Conductance as a function of energy $E$ and Zeeman splitting ${J}$ at finite temperature within the intrinsic pairing model. Plotted with $\mu/\Delta=1.2$, $E_{so}/\Delta=1.5$, $Z=3$, and $T/T_c=0.1$. (b) Line cuts of plot in (a) for different values of ${J}$, ranging from ${J}/\Delta=0$ to ${J}/\Delta=3$ in steps of ${J}/\Delta=0.1$. Plots offset by $0.1\times e^2/h$ for clarity.}
\end{figure*}

Finally, we relax all constraints on the parameters of the model and numerically solve for the conductance at finite temperature. In Fig.~\ref{finiteT}, we map out the energy and magnetic field dependence of the conductance for $\mu/\Delta=1.2$, $E_{so}/\Delta=1.5$, $Z=3$, and $T/T_c=0.1$. At zero field, there is a single peak in the conductance spectrum at $E=\Delta$. As the field is increased, the position of this peak is shifted to lower energies, signifying the closing of the nontopological gap. At fields above the critical value ${J}_c=\sqrt{\mu^2+\Delta^2}$, a zero-energy peak emerges due to the Majorana fermion that is present in the topological phase.

For our specific choice of parameters, we find that the amplitude of the zero-bias peak is about $G(0)=1.2\times e^2/h$. While this value is not as low as what has been observed experimentally in similar nanowire systems [which typically range anywhere between $G(0)\sim0.1-0.9\times e^2/h$], \cite{Mourik:2012,Deng:2012,Das:2012,Zhang:2016} our calculation does demonstrate that finite temperature can effectively reduce the zero-bias conductance. Whereas the temperature dependence of the zero-bias conductance is controlled by a single parameter ($Z_\text{eff}$) deep in the topological phase, in the vicinity of the phase transition the temperature dependence is determined by an interplay between all parameters of the problem.

\section{Tunneling model} \label{Sec3}
\subsection{Effective Hamiltonian} \label{EffectiveH}
In Sec.~\ref{Sec2}, the proximity effect was incorporated through a BCS-like pairing term in the Hamiltonian of the nanowire. However, this treatment neglects the presence of the underlying superconductor. In this section, we incorporate the superconducting substrate by describing the proximity effect in the bulk of the superconducting segment of the nanowire with a Hamiltonian of the form
\begin{equation} \label{Hamiltonian2}
H=H_{NW}+H_B+H_{S}+H_t.
\end{equation}
$H_{NW}$ and $H_B$ are given by the Fourier transforms (to momentum space) of Eqs.~(\ref{HNW}) and (\ref{HB}), respectively. The underlying superconductor is described by a BCS Hamiltonian,
\begin{equation} \label{HBCS}
\begin{aligned}
H_{S}&=\sum_\sigma\int\frac{d^3k}{(2\pi)^3}\biggl\{\eta_\sigma^\dagger(\kk)\xi_{kS}\eta_\sigma(\kk) \\
	&+\Delta\left[\eta_\downarrow(\kk)\eta_\uparrow(-\kk)+H.c.\right],
\end{aligned}
\end{equation}
where $\eta_\sigma^\dagger(\kk)$ [$\eta_\sigma(\kk)$] creates (annihilates) a state of spin $\sigma$ and momentum $\kk$ in the superconductor, $\xi_{kS}=k^2/2m_S-\mu_S$ ($m_S$ and $\mu_S$ are the effective mass and Fermi energy of the superconductor), and $\Delta$ is the pairing potential of the superconductor. The proximity effect is incorporated through a term which describes spin- and momentum-conserving \cite{note} tunneling between the superconductor and nanowire,
\begin{equation}
H_t=-t\sum_\sigma\int\frac{d^3k}{(2\pi)^3}\left[\psi_\sigma^\dagger(k_x)\eta_\sigma(\kk)+H.c.\right].
\end{equation}
By integrating out the superconducting degrees of freedom (see Appendix~\ref{Intoutapp} for details), we find an effective theory describing superconductivity induced in the nanowire. \cite{Alicea:2012,Potter:2011,Sau:2010prox,vanHeck:2016} Comparing with the intrinsic pairing model of Sec.~\ref{Sec2}, the superconducting substrate can be incorporated by renormalizing both the energy, $E\to E/\Gamma(E)$, and pairing potential, $\Delta\to\Delta[1/\Gamma(E)-1]$, where
\begin{equation}
\Gamma(E)=\left[1+\frac{\gamma}{-i\sqrt{E^2-\Delta^2}}\right]^{-1}
\end{equation}
and $\gamma=\pi\nu_{2D}t^2$ is an energy scale related to the tunneling strength ($\nu_{2D}=m_S/2\pi$ is the density of states of an effective two-dimensional electron gas). Physically, the quantity $\Gamma(E)$ describes the fraction of time that an electron spends in the nanowire (as opposed to the superconductor). \cite{vanHeck:2016}

We assume that our treatment of the proximity effect, which assumes that the wire is infinite and that the superconductor is unbounded in the plane of the wire, is still applicable to the geometry shown in Fig.~\ref{geometry}. We also neglect any feedback effect that the wire may have on the superconductor. While it has been shown that such feedback effects can drastically affect the spectrum of a finite-sized system, they are negligible in infinitely large systems like the one we are considering here. \cite{Reeg:2017}

In the absence of an external magnetic field, the magnitude of the superconducting gap that is proximity-induced in the nanowire ($E_g$) is determined by the strength of tunneling. The size of the induced gap is determined implicitly by the equation
\begin{equation} \label{gapcondition}
E_g^2/\Gamma^2(E_g)=\Delta^2[1/\Gamma(E_g)-1]^2.
\end{equation}
In the strong-tunneling limit  ($\gamma\gg\Delta$), the induced gap is $E_g=\Delta(1-2\Delta^2/\gamma^2)$; in the weak-tunneling limit ($\gamma\ll\Delta$), the induced gap is $E_g=\gamma$. While $\gamma$ is an important parameter of the tunneling model that is very difficult to control, the authors of Ref.~\onlinecite{vanHeck:2016} pointed out that this quantity is experimentally accessible. By solving Eq.~(\ref{gapcondition}), we can express $\gamma$ in terms of the proximity-induced gap $E_g$ (in the absence of the applied field) and the bulk gap of the superconducting substrate $\Delta$,
\begin{equation}
\gamma=E_g\sqrt{\frac{\Delta+E_g}{\Delta-E_g}}.
\end{equation}
Therefore, by measuring both $E_g$ and $\Delta$, it is possible to experimentally determine the tunneling strength $\gamma$.

Finally, we note that the two models considered in this paper are equivalent at low energies ($E\ll\Delta$) and in the weak-tunneling limit ($\gamma\ll\Delta$). Under these two conditions, $\Gamma(E)=(1+\gamma/\Delta)^{-1}$ and we can approximate $E/\Gamma(E)=E$ and $\Delta[1/\Gamma(E)-1]=\gamma$. Therefore, the electron energy in the nanowire is not renormalized by the superconductor, and the effective pairing potential is independent of energy and equal to the induced excitation gap in the nanowire ($\gamma$).
   
\subsection{Conductance in the tunneling model} \label{ConductanceTunneling}
We now move on to calculate the conductance within the tunneling model. The BdG equation that we look to solve is the same as in Eq.~(\ref{BdG}), with the replacements $E\to E/\Gamma(E)$ and $\Delta\to\Delta[1/\Gamma(E)-1]$. We begin by discussing the conductance in the absence of the external field. While this case was rather trivial within the intrinsic pairing model considered in Sec.~\ref{Zerofield}, it is still instructive to consider this simple limit to illustrate some of the main features of the tunneling model.

\subsubsection{Zero-Field Limit $B_\text{ext}=0$} \label{ZeroFieldTunn}
To solve for the conductance in the tunneling model, we can use our solution from Sec.~\ref{Zerofield}, again assuming that $(\mu+E_{so})\gg\Delta$. There are three energy ranges that must be considered separately. When $E<E_g$, transmission into the superconducting segment of the wire is not allowed because the minimum single-particle excitation is $E_g$. Conversely, electrons can be transmitted into the superconducting segment of the nanowire over the energy range $E_g<E<\Delta$. For energies $E>\Delta$, $\Gamma(E)$ becomes a complex quantity (as opposed to the values $0<\Gamma<1$ it takes for $E<\Delta$),
\begin{equation}
\Gamma(E>\Delta)=\frac{E^2-\Delta^2}{E^2-\Delta^2+\gamma^2}-\frac{i\gamma\sqrt{E^2-\Delta^2}}{E^2-\Delta^2+\gamma^2}.
\end{equation}
A complex $\Gamma(E)$ signifies that single-particle excitations are able to enter the superconducting substrate. Therefore, there is no transmission to the superconducting segment of the nanowire.

\begin{figure}[t!]
\centering
\includegraphics[width=0.8\linewidth]{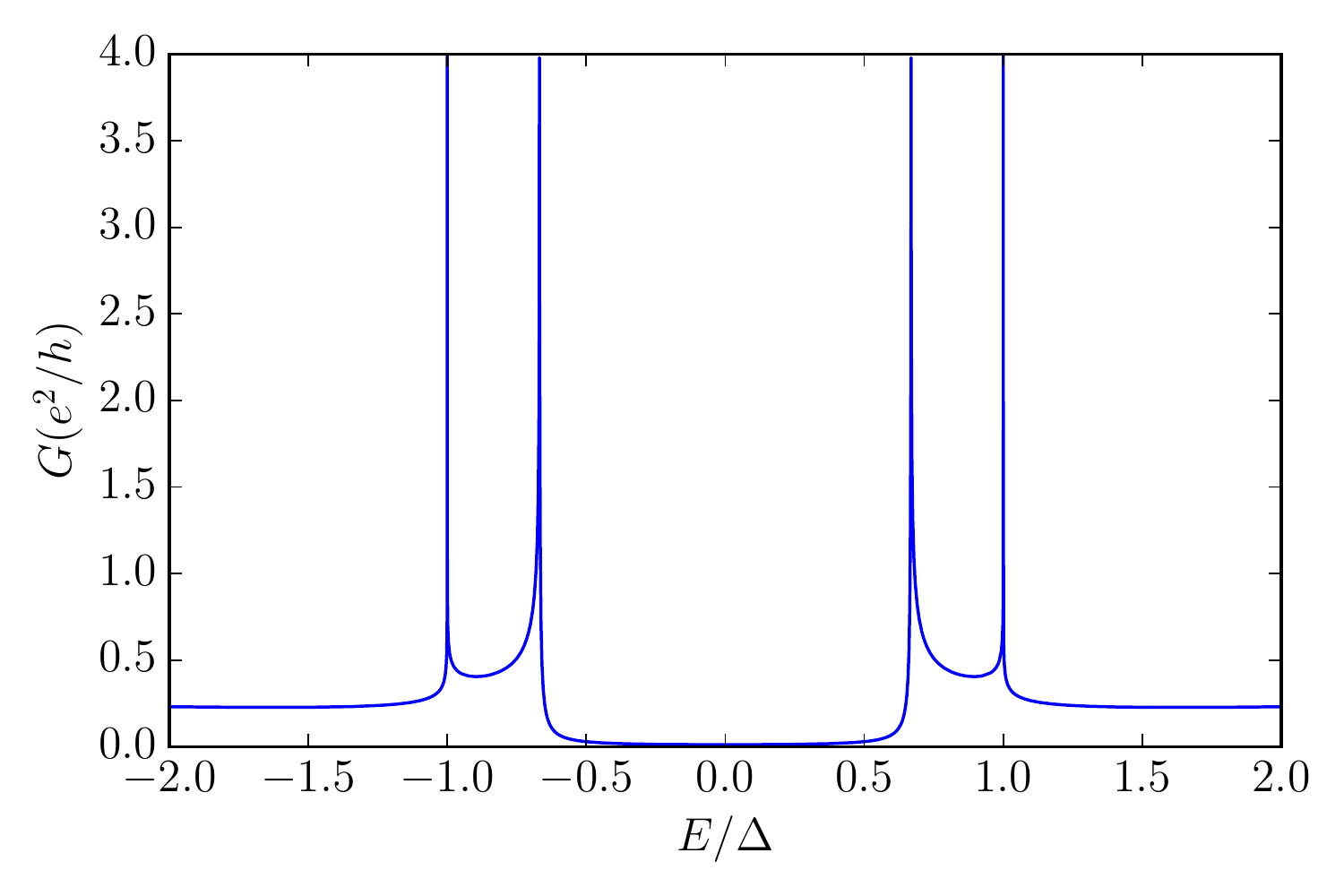}
\caption{\label{nofieldtunn}Conductance obtained within the tunneling model in the limit $(\mu+E_{so})\gg\Delta$ and ${J}=0$ at zero temperature. Plotted for $Z=3$ and $\gamma=1.5$, corresponding to an induced gap of $E_g\approx 0.7\Delta$.}
\end{figure}

The scattering probabilities in the tunneling model are obtained directly from the scattering amplitudes of Eqs.~(\ref{scatteringamps}) by making the appropriate replacements. The probability for an incident spin-up electron to Andreev reflect as a spin-down hole is given by
\begin{equation} \label{Andreevtunn}
|a_{\downarrow\uparrow}|^2=\left\{\begin{array}{cc}
	 \frac{\bar\Delta^2}{\bar\Delta^2(1+2Z^2)^2-4E^2Z^2(1+Z^2)}, & E<E_g \\
	 \frac{\bar\Delta^2}{[E+\sqrt{E^2-\bar\Delta^2}(1+2Z^2)]^2}, & E_g<E<\Delta \\
	 \frac{\Delta^2\gamma^2}{D_1}, & E>\Delta
\end{array}\right.,
\end{equation}
where we define $\bar\Delta=\Delta[1-\Gamma(E)]$, $D_1=E^2(\gamma^2+\Omega^2)+(1+2Z^2)^2|\beta|^2+2E(1+2Z^2)(\gamma\,\text{Re}\beta+\Omega\,\text{Im}\beta)$, $\Omega^2=E^2-\Delta^2$, and $\beta^2=E^2(\gamma+i\Omega)^2-\Delta^2\gamma^2$. Similarly, the normal reflection probability of an incident spin-up electron is given by
\begin{equation} \label{Normaltunn}
|r_{\uparrow\uparrow}|^2=\left\{\begin{array}{cc}
	 \frac{4(\bar\Delta^2-E^2)Z^2(1+Z^2)}{\bar\Delta^2(1+2Z^2)^2-4E^2Z^2(1+Z^2)}, & E<E_g \\
	 \frac{4(E^2-\bar\Delta^2)Z^2(1+Z^2)}{[E+\sqrt{E^2-\bar\Delta^2}(1+2Z^2)]^2}, & E_g<E<\Delta \\
	 \frac{4|\beta|^2Z^2(1+Z^2)}{D_1}, & E>\Delta
\end{array}\right..
\end{equation}
From Eqs.~(\ref{Andreevtunn}) and (\ref{Normaltunn}), we see that
\begin{equation}
|a_{\downarrow\uparrow}|^2+|r_{\uparrow\uparrow}|^2<1
\end{equation}
for energies $E>\Delta$. Because transmission is not allowed for these energies, this means that the scattering probability is not conserved within the tunneling model. Equivalently, the continuity equation is not satisfied in the nanowire, as particles with energies $E>\Delta$ are lost to the superconducting substrate.

\begin{figure}[t!]
\centering
\includegraphics[width=\linewidth]{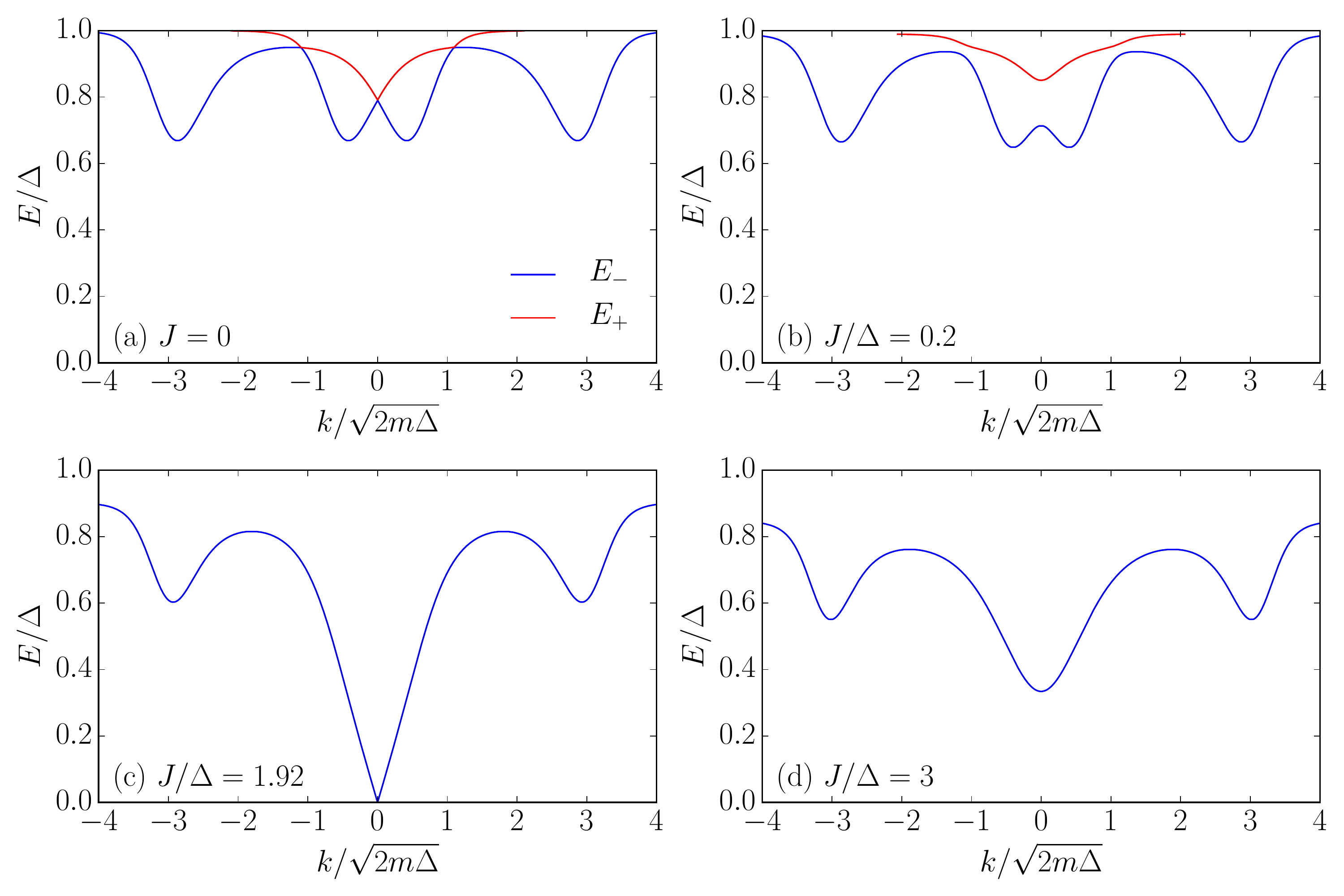}
\caption{\label{spectrumtunn}Bulk excitation spectrum in a nanowire with tunneling-induced superconductivity for various field strengths at zero temperature. (a) In the absence of the field, the lower branch of the spectrum has an excitation gap $E_g$ and the spectrum is degenerate at $k=0$. (b) Application of a field lifts the degeneracy and reduces the gap on the lower branch at $k=0$. (c) At the critical field strength ${J}_c=\sqrt{\mu^2+\gamma^2}$, the gap closes. (d) The gap is reopened in the topological phase as the field is increased beyond ${J}_c$. In all cases, there are no bulk states at energies $E>\Delta-(g_S/g){J}$, as the energy acquires an imaginary part. This signifies that these states are lost to the superconducting substrate. All plots shown for $\mu/\Delta=1.2$, $E_{so}/\Delta=1.5$, $\gamma/\Delta=1.5$, and $g_S/g=0.05$.}
\end{figure}

The conductance is found using Eq.~(\ref{Gnofield}),
\begin{equation} \label{Gtunnnofield}
G(E)=\frac{2e^2}{h}\left\{\begin{array}{cc}
	 \frac{2\bar\Delta^2}{\bar\Delta^2(1+2Z^2)^2-4E^2Z^2(1+Z^2)}, & E<E_g \\
	 \frac{2E}{E+\sqrt{E^2-\bar\Delta^2}(1+2Z^2)}, & E_g<E<\Delta \\
	 1+\frac{\Delta^2\gamma^2-4|\beta|^2Z^2(1+Z^2)}{D_1}, & E>\Delta
\end{array}\right..
\end{equation}
The conductance is plotted in Fig.~\ref{nofieldtunn} choosing $Z=3$ and $\gamma=1.5$ (corresponding to $E_g\approx0.7\Delta$). In the tunneling model, there are two distinct peaks in the conductance spectrum at $E=E_g$ and $E=\Delta$, and both peaks are fixed to an amplitude of $4e^2/h$ [this can be seen in Eq.~(\ref{Gtunnnofield})]. These two peaks correspond to the proximity-induced gap of the nanowire and the bulk gap of the superconducting substrate, respectively.

\subsubsection{Finite Magnetic Field and Finite Temperature} \label{FieldTunn}
When an external magnetic field is introduced, the excitation spectrum in the bulk of the superconducting segment of the wire is given implicitly by the equation
\begin{equation} \label{topspectrumtunn}
\begin{aligned}
E_\pm^2&/\Gamma^2(E_\pm)={J}^2+\Delta^2[1/\Gamma(E_\pm)-1]^2+\xi_k^2+(\alpha k)^2 \\
	&\pm2\sqrt{{J}^2\Delta^2[1/\Gamma(E_\pm)-1]^2+{J}^2\xi_k^2+(\alpha k)^2\xi_k^2}.
\end{aligned}
\end{equation}
The topological phase transition can be found by determining when the $k=0$ gap in the spectrum closes, or when $E=0$ solves Eq.~(\ref{topspectrumtunn}). The critical field strength corresponding to the transition is given by \cite{Stanescu:2011}
\begin{equation}
{J}_c=\sqrt{\mu^2+\gamma^2}.
\end{equation}
It is not the induced gap $E_g$ which enters the topological criterion (as in the intrinsic pairing model), but rather the tunneling strength $\gamma$. Therefore, if the coupling between the superconductor and nanowire is made too strong, it will require very large applied fields to reach the topological phase.

\begin{figure}[t!]
\centering
\includegraphics[width=\linewidth]{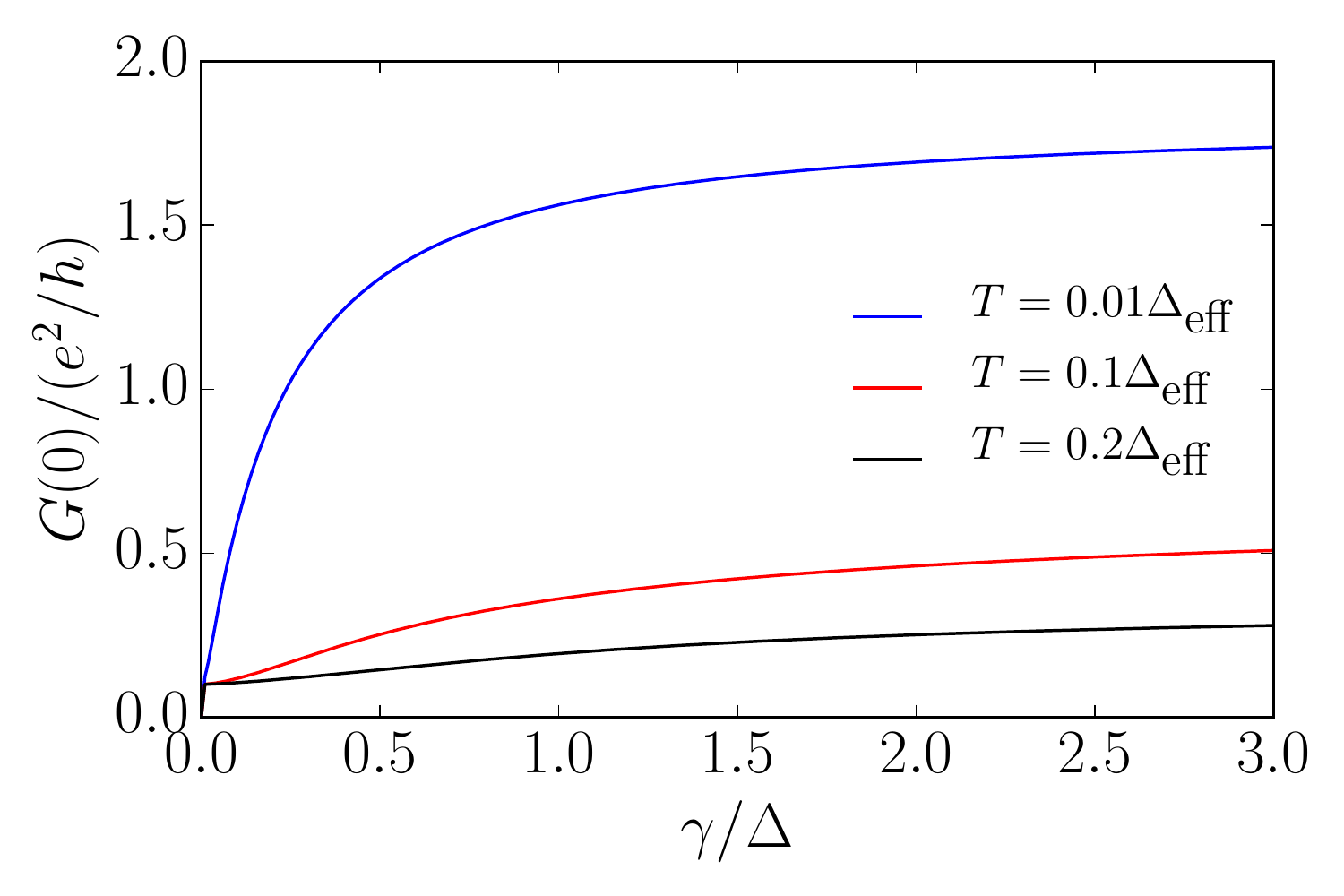}
\caption{\label{G0vsg}Zero-bias conductance deep in the topological phase $J\gg\gamma$, plotted as a function of tunneling strength $\gamma$ (in units of the bulk superconducting gap $\Delta$) for several temperatures $T$ (in units of $\Delta_\text{eff}=\Delta\sqrt{E_{so}/J}$). All curves plotted with $Z_\text{eff}=3$.}
\end{figure}

\begin{figure*}
\centering
\includegraphics[width=0.9\linewidth]{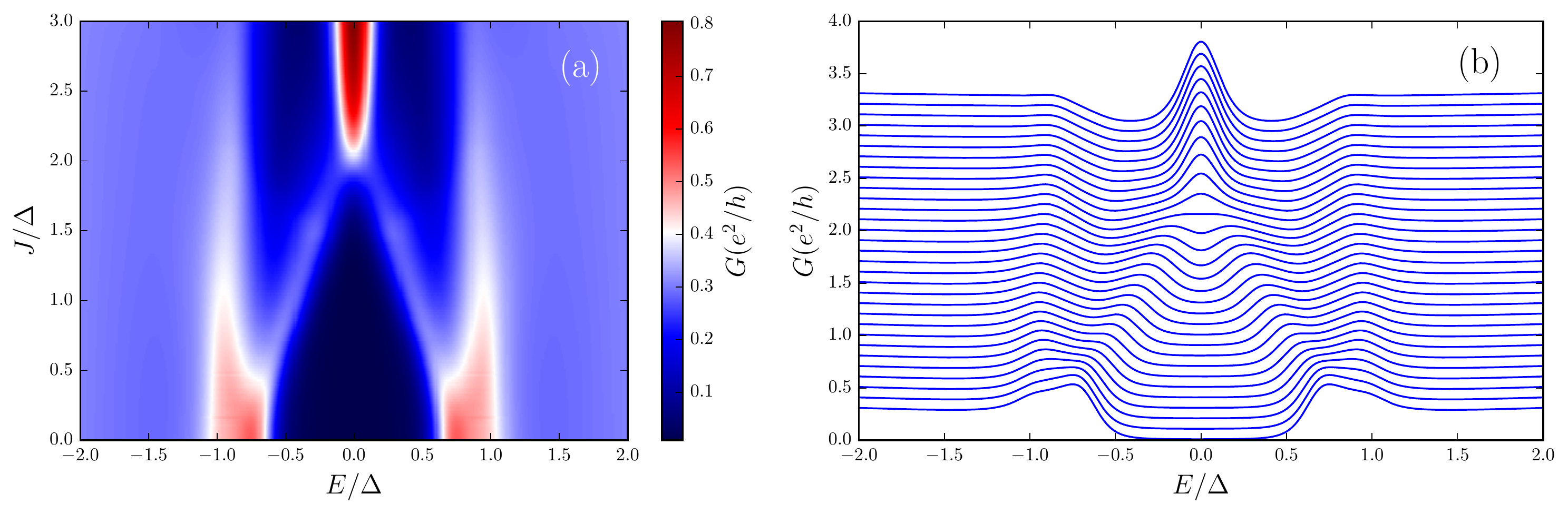}
\caption{\label{finiteTtunn}(a) Conductance as a function of energy $E$ and Zeeman field ${J}$ at finite temperature within the tunneling model. Plotted with $\mu/\Delta=1.2$, $E_{so}/\Delta=1.5$, $\gamma/\Delta=1.5$, $Z=3$, $T/T_c=0.1$, and $g_S/g_N=0.05$. (b) Line cuts of plot in (a) for different values of ${J}$, ranging from ${J}/\Delta=0$ to ${J}/\Delta=3$ in steps of ${J}/\Delta=0.1$. Plots offset by $0.1\times e^2/h$ for clarity.}
\end{figure*}

It is also expected that the external field applied to reach the topological phase in the nanowire will have a detrimental effect on the superconducting substrate. Even for large $g$-factor materials like InSb ($g\sim40$)\cite{Kammhuber:2016}, the field needed to reach the topological phase is $B_\text{ext}\sim1$ T. The Zeeman splitting induced by a magnetic field reduces the excitation gap of an $s$-wave superconductor while leaving the pairing potential $\Delta$ unchanged (provided that the applied field strengths do not reach the Clogston limit \cite{Clogston:1962}). To model the effects of the field, we absorb the Zeeman splitting to define a ``tunneling gap" which depends on both the field strength and the temperature as
\begin{equation}
\Delta({J},T)=\Delta(0,0)\tanh(1.74\sqrt{T/T_c-1})-(g_S/g){J},
\end{equation}
where $g_S=2$ is the Land\'{e} $g$-factor of the superconductor. We neglect the suppression of $\Delta$ due to orbital effects of the field, which is a reasonable assumption if the applied field is much smaller than $H_{c2}$ of the superconductor (as is the case for example in Ref.~\onlinecite{Zhang:2016}, with NbTiN having $H_{c2}>10$ T). Accounting for the suppression of the gap by the Zeeman splitting, the bulk spectrum of the superconducting segment of the nanowire at zero temperature is shown in Fig.~\ref{spectrumtunn}.

The conductance can still be determined using Eqs.~(\ref{Gless}) and (\ref{Gmore}), as the normal segment of the nanowire is unaffected by the superconductor, and our analytic results from Sec.~\ref{Sec2} can be easily extended to the tunneling model by making the replacements $E\to E/\Gamma(E)$ and $\Delta\to\Delta[1/\Gamma(E)-1]$. We now investigate the effect of the tunneling energy $\gamma$ on the zero-bias conductance at finite temperature (i.e., we extend the analytical calculations of Sec.~\ref{Numerics} to the tunneling model). We assume that we are deep in the topological phase ($J\gg\gamma$), and that the tunneling strength $\gamma$ is not so large that the field needed to access this regime destroys superconductivity [$\Delta(J,0)\lesssim\Delta(0,0)$]. At low temperatures $T\ll\Delta,\tilde\Delta_\text{eff}$, where we denote $\Delta=\Delta(J,0)$ and $\tilde\Delta_\text{eff}=\Delta_\text{eff}(\gamma/\Delta)$ ($\Delta_\text{eff}=\Delta\sqrt{E_{so}/J}$ as before, but now $\Delta$ is the gap of the underlying superconductor rather than the gap induced in the wire), and in the tunneling limit $Z_\text{eff}\gg1$ (recall $Z_\text{eff}=Z\sqrt{E_{so}/J}$), the zero-bias conductance is found from Eq.~(\ref{zerobiasfiniteT}) to be
\begin{equation}
G(0)=\frac{2e^2}{h}\int d\varepsilon\left(\frac{1}{\pi}\frac{\tilde W}{\tilde W^2+4\tilde\varepsilon^2}\right)\frac{\pi \tilde W}{4T\cosh^2(\varepsilon/2T)},
\end{equation}
where $\tilde\varepsilon=\varepsilon(1+\gamma/\Delta)$ and $\tilde W=\tilde\Delta_\text{eff}/Z_\text{eff}^2$. Again, for the lowest temperatures $T\ll\tilde W\ll\tilde\Delta_\text{eff}$, the zero-bias conductance is fixed to $2e^2/h$. For higher temperatures $\tilde W\ll T\ll\tilde\Delta_\text{eff}$, we replace the quantity in parentheses by $\delta(\varepsilon)/[2(1+\gamma/\Delta)]$ and the zero-bias conductance is given by
\begin{equation} \label{G0tunnfiniteT}
G(0)=\frac{e^2}{h}\frac{\pi\Delta_\text{eff}}{4TZ_\text{eff}^2}\frac{\gamma}{\Delta+\gamma}\ll\frac{2e^2}{h}
\end{equation}
We see from Eq.~(\ref{G0tunnfiniteT}) that an increased tunneling strength $\gamma$ in turn increases the zero-bias conductance at finite temperature. This is also demonstrated in Fig.~\ref{G0vsg}, where we plot the zero-bias conductance as a function of $\gamma$ for several different temperatures. We obtained Fig.~\ref{G0vsg} by substituting Eq.~(\ref{pwavecond}) (after making the appropriate replacements) into Eq.~(\ref{GT}) and performing the integration numerically. We also find that the zero-bias conductance drops discontinuously to zero for $\gamma=0$; this is due to the fact that at $T=0$, $G(0)=2e^2/h$ if $\gamma>0$ and $G(0)=0$ if $\gamma=0$ (when there is no induced superconductivity in the wire and thus no topological phase).

We now relax all restrictions on the parameter space and numerically calculate the conductance as a function of both energy $E$ and Zeeman field $J$ at finite temperature. The results of our calculation are shown in Fig.~\ref{finiteTtunn}, where we plot $G(E,{J})$ choosing $\mu/\Delta=1.2$, $E_{so}/\Delta=1.5$, $\gamma/\Delta=1.5$, $Z=3$, $g_S/g=0.05$ (or $g=40$), and $T/T_c=0.1$ ($T_c$ here is the critical temperature of the superconducting substrate).

Comparing with the intrinsic pairing model (Fig.~\ref{finiteT}), there are a few qualitative differences. First, the conductance in the tunneling model exhibits two distinct peaks as a function of energy over the entire range of field strengths (as shown in the previous section, these peaks correspond to the induced gap in the nanowire and the bulk gap of the superconductor). The position of the lower-energy peak starts at $E=E_g$ in the absence of the field, decreases as the field is turned on, and is fixed to $E=0$ in the topological phase. The position of the higher-energy peak decreases nearly linearly with the field due to our choice for modeling the field dependence of the BCS gap. Second, as previously discussed, the topological phase transition is shifted to a higher field strength ${J}_c=\sqrt{\mu^2+\gamma^2}$. 

\section{Conclusions} \label{Sec4}
We calculated the conductance of a one-dimensional normal/superconducting nanowire junction within the Blonder-Tinkham-Klapwijk theory in the presence of spin-orbit coupling, an external magnetic field, and conventional superconductivity, utilizing a combination of analytical methods at zero temperature and numerical methods at finite temperature. We directly compared two models of the superconducting proximity effect: one where superconductivity is incorporated through an intrinsic pairing mechanism in the nanowire, and one where superconductivity is induced through a tunnel coupling with a bulk superconducting substrate. We found that the conductance in the tunneling model exhibits an additional peak at the energy corresponding to the gap of the underlying superconductor. While the zero-bias conductance is fixed to $2e^2/h$ in the topological phase at zero temperature (in both models), we showed that finite temperature can significantly reduce the amplitude of the zero-bias peak when the normal and superconducting segments of the nanowire are weakly coupled.

Before concluding, we would like to remark on how our numerical calculations of the conductance at finite temperature within the tunneling model compare with the most recent generation of experiments on InSb nanowires. \cite{Zhang:2016} Choosing realistic parameters for InSb nanowires (as we do in Fig.~\ref{finiteTtunn}), we are able to reproduce most of the qualitative experimental features. These include both the profile of the gap-closing transition as a function of the field and the presence of a secondary peak in the conductance that persists into the topological phase. The largest discrepancy between our calculation and the experiment is that we need to choose a much higher temperature than what is reported ($T_\text{exp}\sim0.01T_c$), as the features produced by our model at lower temperatures are much sharper than those observed. However, as noted in Ref.~\onlinecite{Zhang:2016}, the width of the observed zero-bias peak is larger than what would be expected due solely to thermal broadening. If we incorporate all possible broadening effects (e.g. multiple subbands in the wire, soft tunnel barrier, inelastic and dephasing processes, etc.) into a single effective temperature parameter, then our choice is not so unrealistic.

\acknowledgments
We thank J. Klinovaja, L. Kouwenhoven, D. Loss, and H. Zhang for helpful discussions. This work was supported by the National Science Foundation via grant NSF DMR-1308972 (C.R.R. and D.L.M.), as well as by the Swiss National Science Foundation and the NCCR QSIT (C.R.R.).

\appendix

\section{Boundary Conditions and Scattering Amplitudes for $E_{so}\gg {J},\Delta$} \label{BCapp}
In this appendix, we give explicit forms for both the boundary conditions and scattering amplitudes in the strong spin-orbit limit discussed in Sec.~\ref{Sec2}~\ref{StrongSOC} of the main text. The scattering wave functions in the normal and superconducting segments are given in Eqs.~(\ref{scatteringN}) and (\ref{scatteringS}), respectively, and the boundary conditions that need to be imposed are given in Eqs.~(\ref{BC2}).

First, we consider the case of an incident electron from the lower subband, which corresponds to choosing $\psi_{i-}$ in Eq.~(\ref{scatteringN}). Continuity of the wave function at $x=0$ imposes a set of four conditions given by
{\small
\begin{subequations} \label{BCapp1}
\begin{gather}
1+r_{+-}u_{J}=t_{e-}u_\Delta+\frac{1}{\sqrt{2}}(t_{h+}v_{\Delta-{J}}+t_{e+}v_{\Delta+{J}}), \\
r_{+-}v_{J}-r_{--}=t_{h-}v_\Delta-\frac{1}{\sqrt{2}}(t_{h+}u_{\Delta-{J}}-t_{e+}u_{\Delta+{J}}), \\
a_{+-}u_{J}=t_{h-}u_\Delta-\frac{1}{\sqrt{2}}(t_{h+}v_{\Delta-{J}}-t_{e+}v_{\Delta+{J}}), \\
a_{+-}v_{J}+a_{--}=t_{e-}v_\Delta+\frac{1}{\sqrt{2}}(t_{h+}u_{\Delta-{J}}+t_{e+}u_{\Delta+{J}}).
\end{gather}
\end{subequations}}
In the boundary condition imposed on the derivative of the wave function, we neglect terms proportional to $1/\alpha$:
{\small
\begin{subequations} \label{BCapp2}
\begin{gather}
1-t_{e-}u_\Delta=iZ(1+r_{+-}u_{J}), \\
r_{--}+t_{h-}v_\Delta=iZ(r_{+-}v_{J}-r_{--}), \\
t_{h-}u_\Delta=iZa_{+-}u_{J}, \\
a_{--}-t_{e-}v_\Delta=iZ(a_{+-}v_{J}+a_{--}).
\end{gather}
\end{subequations}}
Solutions to Eqs.~(\ref{BCapp1}) and (\ref{BCapp2}) are given below in Eq.~(\ref{BCsols}).

Next, we consider the case where the incident electron is from the upper subband, which corresponds to choosing $\psi_{i+}$ in Eq.~(\ref{scatteringN}). Continuity of the wave function imposes a set of four conditions given by
{\small
\begin{subequations} \label{BCapp3}
\begin{gather}
v_{J}+r_{++}u_{J}=t_{e-}u_\Delta+\frac{1}{\sqrt{2}}(t_{h+}v_{\Delta-{J}}+t_{e+}v_{\Delta+{J}}), \\
u_{J}+r_{++}v_{J}-r_{-+}=t_{h-}v_\Delta-\frac{1}{\sqrt{2}}(t_{h+}u_{\Delta-{J}}-t_{e+}u_{\Delta+{J}}), \\
a_{++}u_{J}=t_{h+}u_\Delta-\frac{1}{\sqrt{2}}(t_{h+}v_{\Delta-{J}}-t_{e+}v_{\Delta+{J}}), \\
a_{++}v_{J}+a_{-+}=t_{e-}v_\Delta+\frac{1}{\sqrt{2}}(t_{h+}u_{\Delta-{J}}+t_{e+}u_{\Delta+{J}}).
\end{gather}
\end{subequations}}
Again neglecting terms proportional to $1/\alpha$, the four conditions imposed on the derivative of the wave function are
{\small
\begin{subequations} \label{BCapp4}
\begin{gather}
-t_{e-}u_\Delta=iZ(v_{J}+r_{++}u_{J}), \\
r_{-+}+t_{h-}v_\Delta=iZ(u_{J}+r_{++}v_{J}-r_{-+}), \\
t_{h-}u_\Delta=iZa_{++}u_{J}, \\
a_{-+}-t_{e-}v_\Delta=iZ(a_{++}v_{J}+a_{-+}).
\end{gather}
\end{subequations}}
Solutions to Eqs.~(\ref{BCapp3}) and (\ref{BCapp4}) are also given below in Eq.~(\ref{BCsols}).

To aid in expressing the solutions for the scattering amplitudes, we define the quantities
\begin{subequations}
\begin{gather}
\eta_{\alpha,\beta}^\pm=u_\alpha u_\beta\pm v_\alpha v_\beta, \\
\xi_{\alpha,\beta}^\pm=u_\alpha v_\beta\pm v_\alpha u_\beta.
\end{gather}
\end{subequations}
With these definitions, the scattering amplitudes can be expressed as

\begin{widetext}
{\small
\begin{subequations} \label{BCsols}
\begin{align}
r_{+-}&=\frac{Z}{2D}\left\{u_{J}(Z+i)\bigl[2v_\Delta^2v_{\Delta+{J}}v_{\Delta-{J}}Z^2+2u_\Delta^2u_{\Delta+{J}}u_{\Delta-{J}}(1+Z^2)-u_\Delta v_\Delta\xi^+_{\Delta+{J},\Delta-{J}}(1+2Z^2)\bigr]\right. \\
	&\hspace*{0.5in}\left.-u_\Delta^2v_{J}\xi^-_{\Delta+{J},\Delta-{J}}(Z-i)\right\}, \nonumber \\
r_{--}&=\frac{Z^2}{2D}\left\{2u_{J}v_{J}\eta^-_{\Delta,\Delta-{J}}\eta^-_{\Delta,\Delta+{J}}-\xi_{\Delta,{J}}^+\xi_{\Delta,{J}}^-\xi_{\Delta+{J},\Delta-{J}}^-\right\}, \\
a_{+-}&=\frac{u_\Delta Z}{2D}\left\{u_\Delta v_{J}\xi^+_{\Delta+{J},\Delta-{J}}(Z+i)-u_{J}v_\Delta\xi^-_{\Delta+{J},\Delta-{J}}(Z-i)-2v_{J}v_\Delta v_{\Delta+{J}}v_{\Delta-{J}}(Z+i)\right\}, \\
a_{--}&=\frac{1}{2D}\left\{2u_\Delta u_{J}v_\Delta v_{J}\xi^-_{\Delta+{J},\Delta-{J}}(Z^2-1)+\xi^+_{\Delta,{J}}\xi^-_{\Delta,{J}}\xi^+_{\Delta+{J},\Delta-{J}}Z^2\right. \\
	&\hspace*{0.5in}\left.-2u_\Delta v_\Delta(u_{J}^2u_{\Delta+{J}}u_{\Delta-{J}}-v_{J}^2v_{\Delta+{J}}v_{\Delta-{J}})(1+Z^2)\right\}, \nonumber \\
r_{++}&=\frac{1}{2D}\left\{2u_{J}v_{J}\bigl(v_\Delta^2v_{\Delta+{J}}v_{\Delta-{J}}Z^4-u_\Delta v_\Delta\xi^+_{\Delta+{J},\Delta-{J}}Z^2(1+Z^2)+u_\Delta^2[u_{\Delta+{J}}u_{\Delta-{J}}(1+2Z^2)^2-v_{\Delta+{J}}v_{\Delta-{J}}]\bigr)\right. \\
	&\hspace*{0.5in}\left.-u_\Delta^2v_{J}^2\xi^-_{\Delta+{J},\Delta-{J}}(Z-i)^2-u_{J}^2u_\Delta^2\xi^-_{\Delta+{J},\Delta-{J}}(Z+i)^2\right\}, \nonumber \\
r_{-+}&=\frac{Z(u_{J}^2-v_{J}^2)}{2D}\left\{u_\Delta^2v_{J}\xi_{\Delta+{J},\Delta-{J}}^-(Z-i)-u_{J}(Z+i)\bigl[2v_\Delta^2v_{\Delta+{J}}v_{\Delta-{J}}Z^2+2u_\Delta^2u_{\Delta+{J}}u_{\Delta-{J}}(1+Z^2)\right. \\
	&\hspace*{0.5in}\left.-u_\Delta v_\Delta\xi^+_{\Delta+{J},\Delta-{J}}(1+2Z^2)\bigr]\right\}, \nonumber \\
a_{++}&=\frac{u_\Delta(u_{J}^2-v_{J}^2)}{2D}\left\{2v_\Delta v_{\Delta+{J}}v_{\Delta-{J}}Z^2-u_\Delta\xi^+_{\Delta+{J},\Delta-{J}}(1+Z^2)\right\}, \\
a_{-+}&=\frac{u_\Delta(u_{J}^2-v_{J}^2)Z}{2D}\left\{u_\Delta v_{J}\xi^+_{\Delta+{J},\Delta-{J}}(Z-i)-u_{J}v_\Delta\xi^-_{\Delta+{J},\Delta-{J}}(Z+i)-2v_\Delta v_{J}v_{\Delta+{J}}v_{\Delta-{J}}(Z-i)\right\} \\
D&=u_\Delta^2v_{J}^2v_{\Delta+{J}}v_{\Delta-{J}}+u_\Delta^2u_{J}v_{J}\xi^-_{\Delta+{J},\Delta-{J}}(Z^2-1)-u_{J}^2[\eta^-_{\Delta,\Delta-{J}}Z^2+u_\Delta u_{\Delta-{J}}][\eta^-_{\Delta,\Delta+{J}}Z^2+u_\Delta u_{\Delta+{J}}].
\end{align}
\end{subequations}}
\end{widetext}

\section{Conductance of Spinless Normal Metal/$p$-wave Superconductor Junction} \label{spinlessapp}
In the limit of a strong external magnetic field ${J}\gg E_{so},\Delta$ the nanowire Hamiltonian (\ref{Hamiltonian}) maps onto the low-density limit of the Kitaev model, which is described by Eq.~(\ref{Kitaev}). The BdG equation describing spinless $p$-wave superconductivity is given by
\begin{equation} \label{spinlessBdG}
\left(\begin{array}{cc}
	H_0 & -i\Delta(x)(\partial_x/k_F) \\
	-i\Delta^*(x)(\partial_x/k_F) & -H_0 \end{array}\right)\psi(x)=E\psi(x),
\end{equation}
where $H_0=-\partial_x^2/2m-\mu$, $\Delta(x)=\Delta\theta(x)$, and $k_F=\sqrt{2m\mu}$. Deep in the topological phase the chemical potential satisfies $\mu\gg|\Delta|$. Solving Eq.~(\ref{spinlessBdG}) in the normal segment gives a scattering wave function
\begin{equation}
\psi_N(x)=\left(\begin{array}{c} 1 \\ 0 \end{array}\right)e^{ik_Fx}+r\left(\begin{array}{c} 1 \\ 0 \end{array}\right)e^{-ik_Fx}+a\left(\begin{array}{c} 0 \\ 1 \end{array}\right)e^{ik_Fx},
\end{equation}
where $r$ and $a$ are normal and Andreev reflection amplitudes, respectively. In the semiclassical limit, the scattering wave function in the superconducting segment is given by
\begin{equation} \label{scatteringSspinless}
\psi_S(x)=t_1\left(\begin{array}{c} u_\Delta \\ v_\Delta \end{array}\right)e^{ip_+x}+t_2\left(\begin{array}{c} -v_\Delta \\ u_\Delta \end{array}\right)e^{-ip_-x},
\end{equation}
where $u(v)_\Delta^2=(1\pm\Omega/E)/2$, $\Omega=\sqrt{E^2-|\Delta|^2}$, and $p_\pm=k_F\pm \Omega/v_F$ ($v_F=k_F/m$ is the Fermi velocity). Comparing with the scattering wave function of a conventional $s$-wave superconductor [e.g., the spin-up channel (first and last terms) of Eq.~(\ref{scatteringSRashba})], the only difference in the $p$-wave case is the sign of the upper component of the transmitted hole-like state [the second term of Eq.~(\ref{scatteringSspinless})]. 

In the semiclassical limit, the boundary conditions that must be imposed at $x=0$ are $\psi_S(0)=\psi_N(0)$ and $\partial_x\psi_S(0)-\partial_x\psi_N(0)=2k_FZ\psi(0)$, where $Z=U/v_F$ is a dimensionless barrier strength. Solving, we obtain the scattering amplitudes
\begin{subequations} \label{appBamps}
\begin{gather}
a=\frac{u_\Delta v_\Delta}{u_\Delta^2+Z^2}, \\
r=-\frac{Z(i+Z)}{u_\Delta^2+Z^2}, \\
t_1=\frac{u_\Delta(1-iZ)}{u_\Delta^2+Z^2}, \\
t_2=\frac{iv_\Delta Z}{u_\Delta^2+Z^2}.
\end{gather}
\end{subequations}
Comparing with Eqs.~(\ref{scatteringamps}), we simply replace $(u_\Delta^2-v_\Delta^2)\to(u_\Delta^2+v_\Delta^2)=1$, a direct consequence of the sign difference discussed in the previous paragraph. Given the scattering amplitudes of Eqs.~(\ref{appBamps}), we find a conductance
\begin{subequations}
\begin{align}
G(E<|\Delta|)&=\frac{e^2}{h}\frac{2|\Delta|^2}{|\Delta|^2+4E^2Z^2(1+Z^2)}, \\
G(E>|\Delta|)&=\frac{e^2}{h}\frac{2E\bigl[E+\Omega(1+2Z^2)\bigr]}{\bigl[E(1+2Z^2)+\Omega\bigr]^2}.
\end{align}
\end{subequations}
Note that at $E=0$, the conductance is $G(0)=2e^2/h$ regardless of barrier strength. For finite $Z$, we obtain a peak in the conductance spectrum at zero energy, with the width of this peak determined by $Z$.

\section{Integrating out Superconductor} \label{Intoutapp}
In this appendix, we review the method of integrating out the superconducting degrees of freedom to obtain an effective theory describing a tunnel-coupled nanowire. \cite{Alicea:2012,Potter:2011,Sau:2010prox,vanHeck:2016} We begin with the Hamiltonian described by Eq.~(\ref{Hamiltonian2}), where the nanowire Hamiltonian can be expressed in momentum space as
\begin{equation}
\begin{aligned}
H_{NW}+H_B&=-\frac{1}{2}\sum_{\sigma,\sigma'}\int\frac{dk_x}{2\pi}\biggl[\psi_\sigma(k_x)\mathcal{H}_{\sigma\sigma'}^T(k_x)\psi_\sigma^\dagger(k_x) \\
	&-\psi_\sigma^\dagger(-k_x)\mathcal{H}_{\sigma\sigma'}(-k_x)\psi_{\sigma'}(-k_x)\biggr],
\end{aligned}
\end{equation}
where $\hat{\mathcal{H}}(k_x)=k_x^2/2m-\mu-\alpha k_x\hat\sigma_z-J\hat\sigma_x$. Defining a spinor of second-quantized operators in the Heisenberg representation, $\eta(\kk,\omega)=[\eta^\dagger_\uparrow(\kk,\omega),\eta^\dagger_\downarrow(\kk,\omega),\eta_\uparrow(-\kk,-\omega),\eta_\downarrow(-\kk,-\omega)]^T$, where $\omega$ is a Matsubara frequency, we can express the action of the superconductor in matrix form as
\begin{equation} \label{Ss}
S_S=\frac{1}{2}\int\frac{d\omega}{2\pi}\int\frac{d^3k}{(2\pi)^3}\eta^\dagger(\kk,\omega)\mathcal{S}_S(k_y,\omega)\eta(\kk,\omega).
\end{equation}
In Eq.~(\ref{Ss}), $\mathcal{S}_S(k_y,\omega)=-i\omega+\xi_{kS}\hat\tau_z-\Delta\hat\tau_y\hat\sigma_y$. If we define an additional spinor $\nu(k_x,\omega)=-(t/2)[\psi^\dagger_\uparrow(k_x,\omega),\psi^\dagger_\downarrow(k_x,\omega),\psi_\uparrow(-k_x,-\omega),\psi_\downarrow(-k_x,-\omega)]^T$, we can express the tunneling action as
\begin{equation}
S_t=\int\frac{d\omega}{2\pi}\int\frac{d^3k}{(2\pi)^3}\left[\nu^\dagger(k_x,\omega)\eta(\kk,\omega)+H.c.\right].
\end{equation}
The coherent state path integral for the partition function of the system is then given by
\begin{equation} \label{partition}
\mathcal{Z}=\int D[\bar\psi,\psi]\int D[\bar\eta,\eta]e^{-S[\bar\psi,\psi,\bar\eta,\eta]},
\end{equation}
where $\bar\psi,\psi$ and $\bar\eta,\eta$ are the Grassman variables corresponding to the nanowire and superconductor fermion operators, respectively. Because the action in Eq.~(\ref{partition}) is quadratic, we can integrate out the $\eta$ fermions exactly. Upon doing so, we obtain an effective action describing the nanowire given by $S_\text{eff}[\bar\psi,\psi]=S_{NW}[\bar\psi,\psi]+S_B[\bar\psi,\psi]+\delta S[\bar\psi,\psi]$, where
\begin{equation} \label{deltaS}
\begin{aligned}
\delta S[\bar\psi,\psi]&=t^2\sum_\sigma\int\frac{d\omega}{2\pi}\int\frac{d^3k}{(2\pi)^3}\biggl\{\frac{i\omega+\xi_{kS}}{\omega^2+\xi_{kS}^2+\Delta^2} \\
	&\times\bar\psi_\sigma(k_x,\omega)\psi_\sigma(k_x,\omega)-\frac{\Delta}{\omega^2+\xi_{kS}^2+\Delta^2} \\
	&\times\bigl[\psi_\downarrow(k_x,\omega)\psi_{\uparrow}(-k_x,-\omega)+H.c.\bigr]\biggr\}.
\end{aligned}
\end{equation}
Now all that remains is to carry out the integration over $\kk_\parallel=(k_y,k_z)$. For example, we can evaluate the integral
\begin{equation} \label{integral}
\int\frac{d^2k_\parallel}{(2\pi)^2}\frac{t^2}{\omega^2+\frac{1}{4m_S^2}\bigl[k_x^2+k_\parallel^2-k_{FS}^2\bigr]^2+\Delta^2}=\frac{\gamma}{\sqrt{\Delta^2+\omega^2}},
\end{equation}
where we define an energy scale associated with the tunneling strength $\gamma=\pi\nu_{2D}t^2$, and $\nu_{2D}=m_S/2\pi$ is the two-dimensional density of states. Performing the integration over $\kk_\parallel$ in Eq.~(\ref{deltaS}), we obtain the effective action describing the nanowire,
\begin{equation}
\begin{aligned}
S_\text{eff}[\bar\psi,\psi]&=\sum_{\sigma,\sigma'}\int\frac{d\omega}{2\pi}\int\frac{dk_x}{2\pi}\biggl\{\bar\psi_\sigma(k_x,\omega) \\
	&\times\bigl[i\omega/\Gamma(\omega)-\mathcal{H}_{\sigma\sigma'}(k_x)\bigr]\psi_{\sigma'}(k_x,\omega) \\
	&-\Delta[1/\Gamma(\omega)-1]\bigl[\psi_\downarrow(k_x,\omega)\psi_\uparrow(-k_x,-\omega)+H.c.\bigr]\biggr\},
\end{aligned}
\end{equation}
where $\Gamma(\omega)=\bigl[1+\gamma/\sqrt{\Delta^2+\omega^2}\bigr]^{-1}$. Comparing with the action of a conventional BCS superconductor, we simply need to make the replacements $\omega\to\omega/\Gamma(\omega)$ and $\Delta\to\Delta[1/\Gamma(\omega)-1]$ to describe the superconductivity that is proximity-induced in the nanowire.

\bibliography{bibMajoranaConductance_v2}

\begin{thebibliography}{60}
\bibitem [{not()}]{note}%
  \BibitemOpen
  \href@noop {} {}\bibinfo {note} {If momentum is not conserved by tunneling,
  the results are not changed qualitatively. In this case, the integral in
  Eq.~(\ref{integral}) must be performed over a 3D momentum (this also changes
  the dimensions of $t$ to ensure that the integral is dimensionless). All
  results that we derive can be extended to the non-momentum-conserving case by
  redefining $\gamma=k_{FS}\nu_{2D}t^2$ (rather than
  $\gamma=\pi\nu_{2D}t^2$).}\BibitemShut {Stop}%
\end{thebibliography}

\end{document}